\documentclass[useAMS,usenatbib]{mn2e}
\usepackage{graphicx}
\newcommand{\pr}{\ensuremath{P_{\rm r}}}

\title[The relationship between geometric albedo and polarimetric properties of asteroids]
{On the calibration of the relation between geometric albedo and polarimetric properties 
for the asteroids\thanks{Partly based on observations carried out at 
the Complejo Astron\'omico El Leoncito, operated under agreement between the Consejo Nacional 
de Investigaciones Cient\'{\i}ficas y T\'ecnicas de la Rep\'ublica Argentina and the 
National Universities of La Plata, C\'ordoba, and San Juan.}}
\author[A. Cellino et al.]{A.\ Cellino$^{1}$, S.\ Bagnulo$^{2}$, 
R.\ Gil-Hutton$^{3}$, P.\ Tanga$^{4}$, M.\ Ca\~nada-Assandri$^{3}$, 
\newauthor
and E. F. Tedesco$^{5}$\\
$^{1}$INAF - Osservatorio Astrofisico di Torino, I-10025 Pino Torinese, Italy. 
{\rm E-mail: cellino@oato.inaf.it}\\
$^{2}$Armagh Observatory, College Hill, Armagh BT61 9DG, UK. {\rm E-mail: sba@arm.ac.uk}\\
$^{3}$CASLEO and San Juan National University, San Juan, Argentina. {\rm E-mail: rgilhutton@casleo.gov.ar, 
micanada03@casleo.gov.ar}\\
$^{4}$Observatoire de la C\^ote d'Azur, Nice, France. {\rm E-mail: Paolo.Tanga@oca.eu}\\
$^{5}$Planetary Science Institute, Tucson, USA. {\rm E-mail: eft@psi.edu}
}

\begin{document}

\date{Accepted 2015 May 22.  Received 2015 May 20; in original form 2014 October 15}
\pagerange{\pageref{firstpage}--\pageref{lastpage}} \pubyear{2014}

\maketitle

\label{firstpage}

\begin{abstract}
We present a new extensive analysis of the old problem of finding a
satisfactory calibration of the relation between the geometric albedo
and some measurable polarization properties of the asteroids. To
achieve our goals, we use all polarimetric data at our disposal. 
For the purposes of calibration, we use a limited sample
of objects for which we can be confident to know the albedo with good
accuracy, according to previous investigations of other authors. We
find a new set of updated calibration coefficients for the classical
slope - albedo relation, but we generalize our analysis and we
consider also alternative possibilities, including the use of 
other polarimetric parameters, one being proposed here for the first 
time, and the possibility to exclude from best-fit analyzes the asteroids 
having low albedos. We also consider a possible parabolic fit of the 
whole set of data. 
\end{abstract}


\section{Introduction}\label{intro}
The geometric albedo of a planetary body illuminated by the Sun is the
ratio of its brightness observed at zero phase angle (i.e., measured
in conditions of ideal solar opposition)\footnote{The phase angle is
  the angle between the directions to the Sun and to the observer as
  seen from the object} to that of an idealized flat,
Lambertian\footnote{That is, a surface having a luminous intensity
  directly proportional to the cosine of the angle between the
  observer's line of sight and the surface normal (emission angle). 
  A Lambertian surface exhibits a uniform radiance when viewed from 
  any angle, because the projection of any given emitting area is 
  also proportional to the cosine of the emission angle.} 
disk having the same cross-section. The word ``albedo'' comes from the
Latin word {\em Albus}, which means ``white''. According to its
definition, therefore, the geometric albedo is the parameter used to indicate
whether the surface of a given object illuminated by the Sun appears
to be dark or bright. The albedo is wavelength dependent. In
planetary science, the geometric albedo has been traditionally
measured in the standard Johnson $V$ band centered around 0.55 $\mu$m,
and it is usually indicated in the literature using the symbol $p_V$.

The geometric albedo is a parameter of primary importance. Being an
optical property of a sunlight-scattering surface, it must depend on
composition, as well as on other properties characterizing the
surface at different size scales, including macro- and micro-texture
and porosity. All these properties are the product of the overall
history of an object's surface, and are determined by the interplay of
phenomena as complex as collisions, local cratering, micro-seismology,
space weathering, thermal phenomena, just to mention a few relevant
processes.  The fine structure and composition of the surface affects
properties of the optical emission which determine the results of many
observing techniques, including photometry and spectroscopy. It is
particularly important in determining the state of polarization of the
scattered sunlight in different illumination conditions, this being
the main subject of this paper.

The geometric albedo is also a fundamental parameter when one wants to
determine the size of a small Solar System body, having at disposal
photometric measurements at visible wavelengths. In particular, a
measurement of brightness in $V$ light is not sufficient to
discriminate between a large, dark object and a small, bright one, if
the albedo is unknown.

Geometric albedo should not be confused with the so-called Bond (or spherical)
albedo. The Bond albedo is the fraction of incident 
sunlight that is scattered in all directions and at all wavelengths. 
The Bond albedo is needed to estimate what fraction of incident
radiation is actually absorbed, and therefore contributes to the
energy balance of the body determining its temperature.
It is possible to define the Bond albedo at any given wavelenght $\lambda$, e.g., 
$A_V$ for that at the V-band of the Johnson UBV system, as
$A_\lambda = q_\lambda \cdot p_\lambda$ \citep{MorrisonLebofsky79}, where $p_\lambda$ is the
geometric albedo at wavelength $\lambda$, and $q_\lambda$ is the so-called phase 
integral, first defined by \citet{Russel16} as the integral of the directionally 
scattered flux, integrated over all directions: 
\begin{displaymath}
q_\lambda = 2 \, \int_0^\pi \Phi(\lambda, \alpha) \sin \alpha d\alpha
\end{displaymath}
where $\Phi(\lambda, \alpha)$ is the disk-integrated brightness of the 
object at phase angle $\alpha$. 
Unfortunately, a determination of the phase integral, which requires in principle 
many measurements of the scattered sunlight at visible wavelengths obtained in 
different illumination conditions, is very hard to achieve, and is seldom available
in practice.  

Asteroid sizes and albedos have been historically determined mostly by
means of measurements of the thermal flux at mid-IR wavelengths (the
so-called thermal radiometry technique), generally using space-based
platforms like the IRAS and, more recently, the WISE
\citep{MasieroWISE} and Akari \citep{Usui13} satellites.  At thermal
IR wavelengths the received flux depends primarily on the size of the
emitting object\footnote{And on the temperature distribution across
its surface, including also a contribution from the fraction of the
body facing the observer but not illuminated by the Sun, when
observing at non-zero phase angle}, and only weakly on the 
albedo. In particular, the Bond albedo determines the fraction of the
incident sunlight which is absorbed by the surface and is
available to raise the temperature of the body. The temperature, in
turn, determines the spectrum of the thermal emission in the IR. Since the
Bond albedo of the asteroids is usually fairly low (in general well below
$30$\%), most of the incident solar flux is actually absorbed by the
body, and the intensity of the thermal flux turns out to be mostly
dependent on the size, whereas the dependence upon relatively small 
differences in albedo is much weaker. Moreover, the computation of the
geometric albedo from the Bond albedo, as mentioned above, would require 
a knowledge of the phase integral, which is essentially unknown in
the vast majority of cases. As a consequence, it is not really possible
to {\em solve} simultaneously for size and albedo in practical 
applications of the thermal radiometry technique. What is normally
done is to derive the size from thermal IR data alone (assuming also that
the objects have spherical shapes), and then 
determine the geometric albedo by using the known relation 
\begin{equation}
\log(D) = 3.1236 -0.2H - 0.5 \log(p_V)
\end{equation} \label{Eqn:DHpv}
where $D$ is the diameter expressed in km (supposing that the object
is spherical), $H$ is the absolute magnitude and $p_V$ is the
geometric albedo. 
To do so properly, at least some $V$ magnitude measurements
obtained during the same apparition\footnote{The
  apparition of an asteroid is the interval of time (several weeks)
  before and after each solar opposition epoch, when the object
  becomes visible to the observers} in which the thermal flux 
of the object is measured, would be needed, in order to
derive from them a reliable value of the absolute magnitude $H$. 
Unfortunately, in the real world no measurements of the $V$ flux 
are really done in thermal radiometry campaigns, and $H$ is directly 
taken from available catalogues. 
In turn, these $H$ values are derived from $V$ magnitude data 
(often of quite poor photometric quality), mostly obtained
in different observing circumstances, and using a photometric model of the
variation of $V$ magnitude as a function of phase angle. As mentioned by
several authors, \citep[see, for instance,][]{HG1G2}, the
magnitude-phase relation for asteroids is described by means of
parameters which are generally poorly known, and this introduces further
errors in the geometric albedo determination. 

In summary, it is difficult to obtain very accurate determinations of
the geometric albedo of an asteroid based on thermal radiometry
measurements. Based on the relation described by
Eq.$~$\ref{Eqn:DHpv} the relative uncertainty on the albedo 
should be twice the relative error on the size, in ideal conditions.
In practice, values as high as 50\,\% in geometric
albedo, or even more for small and faint objects, are common even when the
relative error on the size is of the order of $10$\%\footnote{This
can be seen by plotting together for a comparison the albedos found for
many thousands of asteroids observed by both the WISE and Akari
satellites}. The best results require
measurements of the thermal flux to be obtained at different
wavelengths in the thermal IR, an acceptable knowledge of the variation
of $V$ magnitude with phase, and detailed thermo-physical models
which can be developed when a wealth of physical data is available
from different observing techniques (including a knowledge of the shape 
and spin axis orientation).

In this paper, we focus on another possible option to obtain estimates
of the geometric albedo of asteroids, or other atmosphereless solar
system bodies. This is based on measurements of the state of
polarization of the sunlight scattered by the surface in different
illumination conditions, and on the existence of empirical
relations between geometric albedo and polarimetric properties. 
Our present analysis is mainly devoted to
summarize the state of the art of this application of asteroid
polarimetry, and to provide one or more updated forms of the albedo -
polarization relationship, sufficiently accurate to be used in
practical applications of asteroid polarimetry by the largest possible
number of researchers in the future.  We analyze the
current observational evidence taking into account an extensive data set 
available in the literature\footnote{Including the
  polarimetric data available at the NASA \emph{Planetary Data System}
  at the URL address http://pds.jpl.nasa.gov/ (files maintained by
  D.F. Lupishko and I.N. Belskaya), and the data published by
  \citet{RGH11, RGH12, Assandrietal2012}}, including also 
observations carried out mostly at the Complejo
Astronomico El Leoncito (San Juan, Argentina), that have been published 
only recently \citep{Giletal14}. The derivation of the
albedo from polarimetric data is a challenging problem which has been
open for a long time. In the present paper we take into account both
traditional approaches as well as new possible developments suggested
by the data at our disposal.

\section{Asteroid polarimetric data}\label{data}
Classical asteroid polarimetry consists of measurements of the linear
polarization of the light received from asteroids observed at
different phase angles. The observations give directly the degree of
linear polarization and the position angle of the plane of
polarization. This is usually measured with respect to the orientation
of the direction perpendicular to the scattering plane, namely the
plane containing the Sun, the observer and the target. According to
elementary physical considerations (Fresnel reflection) one should
expect the scattered sunlight emerging from the surface of an
atmosphereless planetary body to be linearly polarized
along the direction perpendicular to the scattering plane.  This
expectation is only partly confirmed by the observations. The
asteroid light at visible wavelengths turns out to be, as expected, in
a state of partial linear polarization, but in different observing
circumstances the plane of linear polarization is found to be either
perpendicular (as expected) or parallel (and this is {\em a priori}
unexpected) to the scattering plane. It is therefore customary in
asteroid polarimetry to express the degree of polarization as the
ratio of the difference of intensity of light beam component
$I_{\perp}$ having the electric vector aligned along the plane
perpendicular to the scattering plane minus the intensity
$I_{\parallel}$ of the component having the electric vector aligned
parallel to that plane, divided by the sum of the two
intensities. This parameter is usually indicated as \pr\ in the
literature and is given by:
\begin{displaymath}
\pr = \frac{(I_{\perp} -  I_{\parallel})}{(I_{\perp} + I_{\parallel})}.
\end{displaymath}

According to its definition, the module of \pr\ is the degree of
linear polarization of the received light as explained in elementary
textbooks in physics (because $I_{\perp}$ and $I_{\parallel}$ are
found to be coincident with $I_{\rm max}$ and $I_{\rm min}$ measured through a
polaroid), but the sign of \pr\ can be either positive or negative,
depending on whether $I_{\rm max}$ corresponds to $I_{\perp}$, as should
be expected based on elementary physics, a situation normally referred
to as ``positive polarization''. When $I_{\rm max}$ is found to correspond
to $I_{\parallel}$, \pr\ becomes negative, and this situation is
called ``negative polarization''.

It is important to note that we will always use the \pr\ parameter
throughout this paper every time we will refer to asteroid
polarimetric measurements. Its value will always be expressed (and
plotted in our figures) in percent (\%). We also note that asteroids
are not strongly polarized objects. The degree of linear polarization
turns out to be usually below $2\%$.

\begin{figure}
\begin{center}
\includegraphics[width=8.9cm]{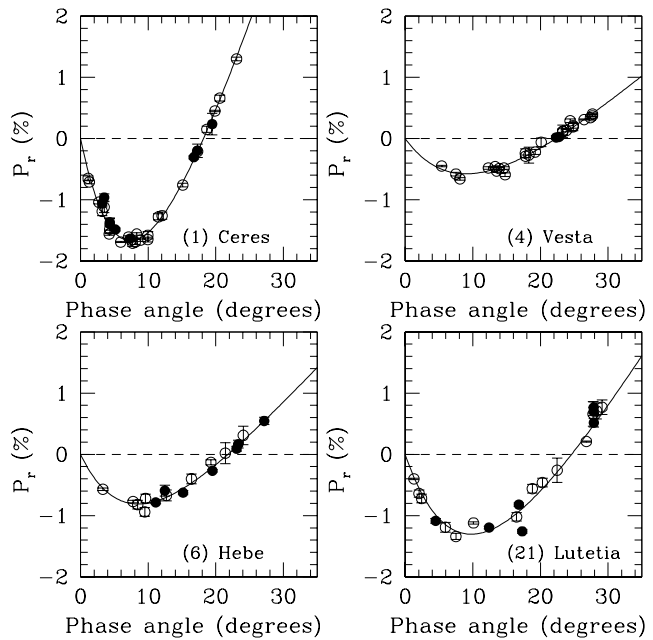}
\end{center}
 \caption{\label{Fig:fig1}
Examples of phase - polarization curves obtained for four large main
belt asteroids belonging to different taxonomic classes: top left: the
dwarf planet (1) Ceres ($G$-class); top right: (4) Vesta ($V$-class);
Bottom left: (6) Hebe ($S$-class); bottom right: (21) Lutetia (for a
long time considered to belong to the old $M$ class, now included in
the $X$ complex). Open symbols refer to data available in the
literature. Full symbols in the plots identify observations obtained
at the CASLEO observatory. Most of them have been published only recently
\citep{Giletal14}. The best-fit lines correspond to the
exponential-linear relation discussed in Sect.~\ref{bestfit}.
} 
\end{figure}
In asteroid polarimetry, by measuring \pr\ at different epochs,
corresponding to different values of the phase angle, it is possible
to obtain the so-called {\em phase - polarization curves}. Some examples are
shown in Fig.~\ref{Fig:fig1}. Many data plotted in this Figure have
been obtained during several observing campaigns carried out at CASLEO
(Complejo AStronomico el LEOncito) in the province of San Juan
(Argentina), using the 2.15 m Sahade telescope \citep{Giletal14}.  
From Fig.~\ref{Fig:fig1}, it is easy to see that asteroids
belonging to very different taxonomic classes tend to exhibit phase -
polarization curves which share in general terms a same kind of
general morphology, but with differences which can be easily seen, and
represent some classical results of asteroid polarimetry 
\citep[see also][]{Penttilaetal2005}.

In particular, all the curves are characterized by a ``negative
polarization branch'', extending over an interval of phase angles
between $0$ degrees up to a value $\alpha_0$ which is commonly found
to be around $20^\circ$ of phase (``inversion angle''). The extreme
value of polarization in the negative branch is traditionally
indicated as $P_{\rm min}$. Around the inversion angle the trend of
\pr\ as a function of phase angle is mostly linear, and the slope of
this linear increase is commonly indicated as $h$. The interval of
phase angles which is accessible to Earth-based observers extends in
the best cases little over $30^\circ$ when observing main belt
asteroids (the possible observing circumstances being determined by
the orbital elements of the objects). The interval of possible phase
angles extends up to much larger values in the case of objects which
can be observed much closer to the Earth, as in the case of many
near-Earth asteroids.

\section{How to calibrate any possible albedo - polarization relation}\label{ST2006}
From Fig.~\ref{Fig:fig1}, it is evident that, at the same phase angle,
different objects exhibit different degrees of polarization. Some
objects are more polarized than others. This can be interpreted in
very general terms as being a consequence of the classical Umov effect
\citep{Umov1905}, which states that the degree of polarization tends
to be inversely proportional to the albedo, according also to
laboratory experiments.

The task of determining the albedo based on available
phase-polarization curves is important in asteroid science.  The
foundations have been laid long ago in some classical papers, like
\citet{ZellnerGradie76} and \citet{Zellneretal77}. Since then, several
authors have tackled the same problem, with analyzes based on new sets
of observations and/or laboratory experiments. The idea is to
determine some suitable relation between the distinctive features of
the phase - polarization curves of some selected sources, and their
albedo, assumed to be known \emph{a priori} with good accuracy. The
set of calibration objects must also be representative of the whole
population.

We cannot assume \emph{a priori} that all asteroids (which are a quite
heterogeneous population), must respect a unique relation between
albedo and polarization properties. Only \emph{a posteriori} can we
assess whether it is possible to accurately derive the albedo using a
unique relationship independent on the taxonomic class.  The main task
is therefore to find a suitable representation of such a relation
between albedo and polarization, to be tested over the maximum
possible number of calibration objects belonging to different
taxonomic classes.

In this respect, one must analyze available polarimetric data for the
largest possible sample of objects having a well-known albedo, in
order to find evidence of some general and satisfactorily accurate
relation between albedo and polarimetric properties.  Unfortunately,
it is not so easy to implement this simple approach in practical
terms.  The use of laboratory experiments, based in particular on
polarimetric measurements of meteorite samples, for calibration
purposes seems to be natural, and was adopted in the 70s, but it is
not exempt from problems.  There are some technical difficulties,
including the need of observing the specimens at zero phase angle to
measure their albedos, something which is in general not a trivial
task. In a classical paper, \citet{Zellneretal77} also mentioned some
problems in correctly assessing the instrumental polarization in the
lab.  Another general problem is to ensure that the used meteorite
samples are really representative of the behavior of asteroids as they
appear to remote observers. Unfortunately, the meteorite samples have
to be treated to reproduce the texture of their asteroid parent
bodies. For instance, \citet{Zellneretal77} noted that to find
similarities between the phase-polarization curves of asteroids and
meteorite samples, the latter have to be first crushed, and the way to
do that has consequences on the derived polarization
properties. Moreover, the authors noticed how difficult it is to
eliminate from meteorite samples any source of terrestrial,
post-impact alteration. For all these reasons, starting from the 90s
most attempts of calibration of the albedo - polarization relation
have been based on the direct use of asteroid albedo values obtained
from other techniques of remote observation.

Several papers have been based on the idea of using for calibration
purposes some sets of asteroids for which the geometric albedo had been derived
from thermal radiometry observations, mostly consisting of old IRAS
measurements, or, more recently, WISE data. This approach has the
advantage of being able to use for calibration many objects belonging
to practically all known taxonomic classes. There are, however,
several problems.  First, albedos derived from thermal IR data are
model-dependent. They depend on the choice of some parameters, which
are needed to simulate the distribution of temperature on the asteroid
surface, and the dependence of the irradiated thermal flux in
different directions. Apart from a limited number of cases, albedo
values determined by thermal radiometry data, particularly for objects
for which we have little information coming from other sources, are
simply too inaccurate for the purposes of a robust
calibration. This is a consequence of the problems discussed in
Section \ref{intro} concerning the general lack of simultaneous
photometric data at visible wavelengths, and the consequent use of
values of absolute magnitude that are affected by large errors. To add
some confusion, in the past the catalog of IRAS-based asteroid albedos
changed with time. In particular, the first published catalog of IRAS
albedo values \citep{IMPS} had been built using thermal IR data
coupled with estimated absolute magnitudes computed prior the
introduction of the ($H$, $G$) asteroid photometric system. Using
these data, \citet{LupMoh} derived a first set of values for the
calibration parameters included in the so-called \emph{slope-albedo}
law (see Section \ref{classical}).  Subsequently, a new IRAS albedo
catalog was produced using absolute magnitudes expressed in the ($H$,
$G$) system \citep{SIMPS}, and these values were used by
\citet{Cellinoetal99} to derive an alternative calibration of the
slope-albedo law. This problem should be expected to arise again,
due to the fact that IAU has recently recommended the use of a
new photometric system \citep[$H, G_1, G_2$, see][]{HG1G2}, implying 
that the albedo catalogues obtained using IRAS, and more recently, 
WISE data, should be updated again.

Currently, different authors use different calibrations available in
the literature, including, in addition to those just mentioned above,
also much older calibrations by \citet{Zellneretal74} and
\citet{Zellneretal77}, which were mainly based on laboratory
experiments. This is not an ideal situation, and one of the major
goals of this paper is just to provide one or more updated forms of
the albedo - polarization relationship, to be used by most researchers
in the future, depending on the polarimetric data at their disposal.

\citet{Masiero12} proposed a new kind of calibration based on
different polarimetric parameters and using a sample of $177$
asteroids for which the albedo has been estimated from WISE thermal
radiometry data. In many cases, the objects were observed by WISE at
fairly large phase angles, and the problem of assigning in these cases
reliable values of corresponding magnitude in the visible is particularly 
difficult.  As a consequence, in this
paper we will also make a new test of the \citet{Masiero12} approach,
but using a different sample of objects having albedos not derived
from thermal radiometry data (see Section \ref{altapp}).

\citet{ShevTed} proposed to use for calibration
purposes a limited sample of asteroids for which both the size is
known with extremely high accuracy, and also the absolute magnitude in
the visible is well known, being based on large data sets of available
photometric data.  As for the size, the most accurate values are
certainly those obtained either \emph{in situ} by space probes, or
those obtained by accurate observations of stellar occultations. At
least for some of these objects, also the absolute magnitude values
listed in the catalogs can be reasonably reliable, although we should
always remember that the absolute magnitude is not, strictly speaking,
a fixed parameter, but it varies at different epochs, being dependent
on the varying aspect angle of the object. The aspect angle determines
the extent of the cross-section of the illuminated surface visible by
the observer, and varies at different apparitions.  This variation of visible
cross-section depends on the overall shape of the object (it is zero
for an ideal sphere) and on the orientation of the rotation axis.

Limiting our analysis to the best-observed objects, for which the size
and the absolute magnitude are supposed to be well known, the albedo
$p_V$ can be derived using the relation between size, albedo and
absolute magnitude (Eq.$~$\ref{Eqn:DHpv})

The list of objects with reliable albedo proposed by \citet{ShevTed}
includes $61$ objects. Among them, there are some of the largest and
most observed asteroids, including (1) Ceres, (2) Pallas, (3) Juno and
(4) Vesta. We note that in the case of (4) Vesta, however, we use
a different value of albedo, namely $0.35 \pm 0.02$, based on the most recent, 
and very accurate value of size measured {\em in situ} by the Dawn probe.
The uncertainty in albedo for this asteroid depends on the fact that
the disk-integrated albedo tends to change at different rotation angles
\citep{CellinoDawn}.
Many objects of the \citet{ShevTed} list are much fainter
(including some small targets of space missions) and several have
never been observed in polarimetry. In this paper we follow the
approach indicated by \citet{ShevTed}. This does not mean that we are
not aware of some problems: first, we know that the \citet{ShevTed}
object list is now fairly old and needs an updating. This includes
both considering a larger, currently available set of high-quality
stellar occultation data, as well as using more accurate values of the
absolute magnitude, to be computed according to the new ($H$, $G_1$,
$G_2$) photometric system 
adopted by the International Astronomical Union. We plan to produce an
updated and possibly longer list of calibration targets in the near
future, but we postpone this to a separate paper. In our present work
we lay the foundations for any future analysis taking profit of a
larger list of reliable asteroid albedos. The still limited data base
of asteroid polarimetric observations is for the moment the main
limiting factor for the investigations in this field.

We have long been involved in an observing program of polarimetric
observations of asteroids belonging to the \citet{ShevTed} list, in
order to improve significantly the coverage of the phase-polarization
curves for these objects. So far, we have been able to obtain decent
phase-polarization curves for only a limited sample of the whole list,
taking also into account that some objects would require
the availability of larger telescopes and/or better detectors, as well
as a larger amount of dedicated observing time. The results presented
in this paper, which follow a previous preliminary analysis published
by \citep{Cellinoetal2012}, are already sufficient to find an updated
set of calibration parameters for the classical form of the
slope-albedo law adopted by most authors in the past. The new data
also allow us to explore new possible ways to express the relation
between albedo and polarimetric properties, which will be probably the
future in this field once the data-set of asteroid polarimetric
measurements will grow significantly in a hopefully not-too distant
future.
\begin{figure}
\begin{center}
\includegraphics[width=88mm]{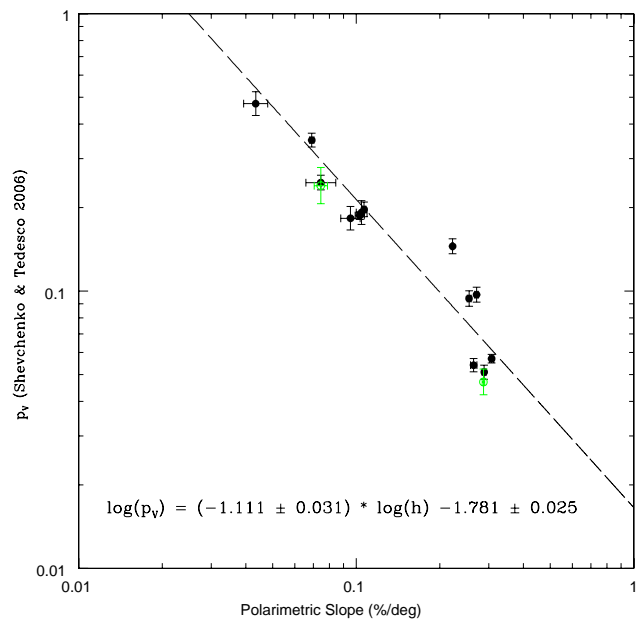}
\end{center}
 \caption{The slope-albedo relation, in log-log scale, for $15$ asteroids of the list of 
\citet{ShevTed} for which a computation of the polarimetric slope is possible based on 
measurements obtained so far. Objects for which we have at disposal at least $10$ polarimetric 
measurements are indicated by full, black symbols. Objects having a number of observations
between $5$ and $10$ are displayed using open, green symbols. The obtained polarimetric slopes 
of all the objects in this plot have been used in the computation of the linear best-fit
that is plotted together with the individual data. The
corresponding values of the $C_1$ and $C_2$ calibration coefficients are also indicated.}
 \label{Fig:slopealbloglogfit}
\end{figure}

\section{The classical approach: the slope-albedo ``law''}\label{classical}

In practically all papers devoted in the past to this subject, the
relation between geometric albedo and polarimetric properties has been
assumed to be one of the following ones:
\begin{equation}
\log(p_V)=C_1 \log(h) + C_2 \label{Eqn:hpv}
\end{equation}
\begin{equation}
\log(p_V)=C_3 \log(P_{\rm min}) + C_4 \label{Eqn:pminpv}
\end{equation}
In Eq.$~$(\ref{Eqn:hpv}), which was originally proposed as early as
in the 70s in the first pioneering investigations by B. Zellner and
coworkers, $h$ is the so-called polarimetric slope, namely the slope
of the linear variation of \pr\ as a function of phase angle,
measured at the inversion angle (see Section \ref{data}). 
In Eq.$~$(\ref{Eqn:pminpv}), the polarimetric parameter is instead $P_{\rm  min}$, 
namely the extreme value of negative polarization.

Most investigations available in the literature \citep[see, for
  instance,][and references therein]{Cellinoetal2012}, have used
Eq.$~$(\ref{Eqn:hpv}), which is normally known as the \emph{slope -
  albedo} relation, or law. In fact, according to many authors,
Eq.$~$\ref{Eqn:pminpv} leads to less (or much less) accurate
results than Eq.$~$\ref{Eqn:hpv}. We will come back to this point
in Section \ref{bestfit}, while in the rest of this Section we will
focus on Eq.$~$\ref{Eqn:hpv}.

The measurement of the polarimetric slope $h$ should be done, in
principle, by measuring the degree of linear polarization \pr\ in a
narrow interval of phase angles surrounding the inversion angle. In
practical terms, however, the observers rarely have at disposal an
ideal coverage of the phase - polarization curve, and often the
polarimetric slope is derived by making a linear fit of a few \pr\
measurements, located not so close to the inversion angle as one would
generally hope. We will see in the next Sections some new possible
approach to derive $h$ when one has at disposal a good coverage of the
phase - polarization curve. For the moment, however, in a first
treatment of available polarimetric data for the objects of the
\citet{ShevTed} list, we adopt the usual techniques, and we derive $h$
from a linear least-squares fit of all available \pr\ data. In
particular, in order to use only homogeneous and high-quality data:
\begin{itemize} 
\item We limit our analysis to polarimetric measurements obtained in the standard $V$ filter.
\item we only use values of linear polarization \pr\ having nominal errors less than $0.2$\%.
\item We use only polarimetric measurements obtained at phase angles larger than
or equal to $14^\circ$ of phase, a value generally well beyond the
phase corresponding to $P_{\rm min}$, and in a region of the negative
polarization branch where \pr\ starts to increase linearly with
phase.
\item We require to have at least $5$ accepted measurements, and that
the interval of phase angles covered by the data is not less than $3^\circ$.
\end{itemize}

A smaller number of measurements, and a narrower interval of covered phase 
angles, would make the adopted polarimetric slopes more uncertain. 
The use of \pr\ data having fairly large error bars, up to 0.2, is
suggested by the general scarcity of polarimetric data, but the effect
of low-quality measurements is mitigated because in the computation of
the best-fit curves, we weight the data according to the inverse of
the square of their associated errors. As for the nominal errors of
the \citet{ShevTed} albedos, which are not explicitly listed by the
authors, we derived them using the quality codes listed in the above
paper, according to their meaning as indicated by the authors.

In this way, we were able to compute reliable polarimetric slopes for $15$
calibration objects. Taking the corresponding albedos from the
\citet{ShevTed} paper, we can plot $h$ as a function of albedo, in a
log-log scale, from which the coefficients $C_1$ and $C_2$ in 
Eq.$~$(\ref{Eqn:hpv}) can be derived using simple least-squares
computations. The nominal errors of both the calibration albedos and 
of the derived polarization slopes were taken into account in the computation.

The results, including also the obtained values of the $C_1$ and
$C_2$ calibration parameters, are shown in 
Fig.$~$\ref{Fig:slopealbloglogfit}. The values of $C_1$ and $C_2$, together
with their corresponding errors, are also given in Table 1.  A few
considerations are suggested by looking at
Fig.$\:$\ref{Fig:slopealbloglogfit}. First, the linear best-fit 
solution seems, as a first approximation, fairly reasonable. 
\begin{figure}
\begin{center}
\includegraphics[width=88mm]{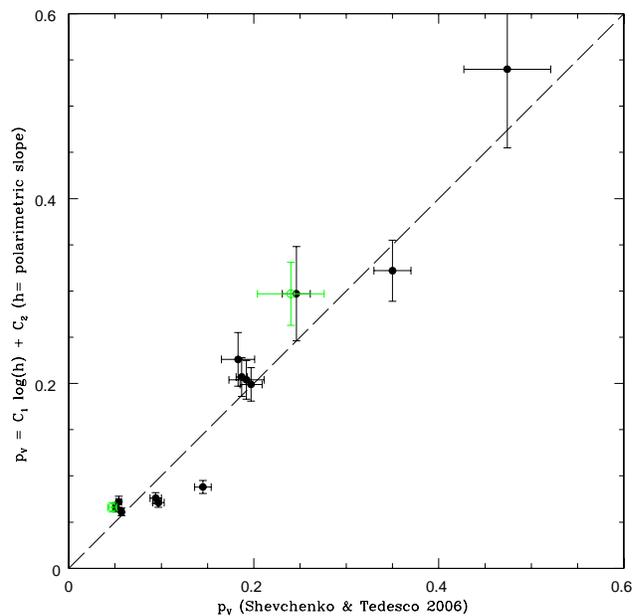}
\end{center}
 \caption{Comparison between the albedos of $15$ objects observed in our campaign, derived using our
 new calibration coefficients of the slope-albedo relation, and the corresponding albedo values 
 given by \citet{ShevTed}. The meaning of the symbols is the same as in 
 Fig.$\:$\ref{Fig:slopealbloglogfit}}
 \label{Fig:albalb}
\end{figure}
\begin{figure}
\begin{center}
\includegraphics[width=88mm]{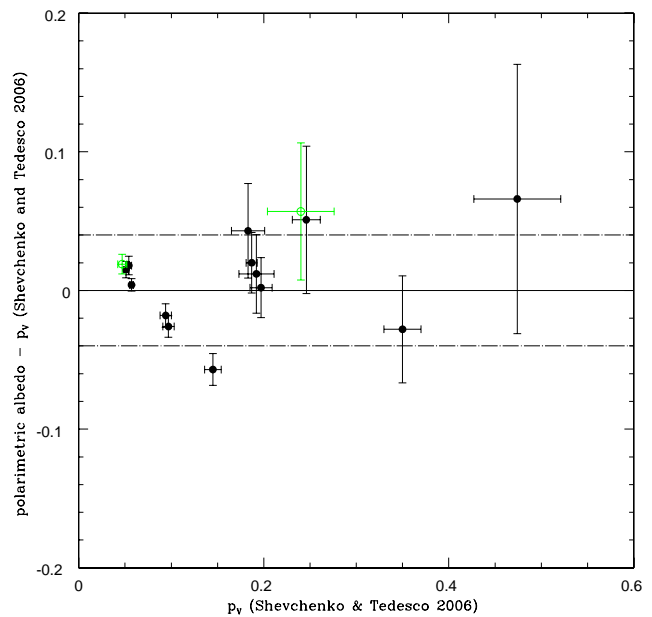}
\end{center}
 \caption{Plot of the difference between the albedo values derived using the new calibration 
 coefficients of the slope-albedo relation for objects included in the \citet{ShevTed} list and the 
 corresponding values found by the above authors. The dashed lines correspond to differences of 
 $\pm 0.04$ in albedo. The meaning of the symbols is the same as in 
 Fig.$\:$\ref{Fig:slopealbloglogfit}.}
 \label{Fig:diffalb}
\end{figure}
\begin{figure}
\begin{center}
\includegraphics[width=88mm]{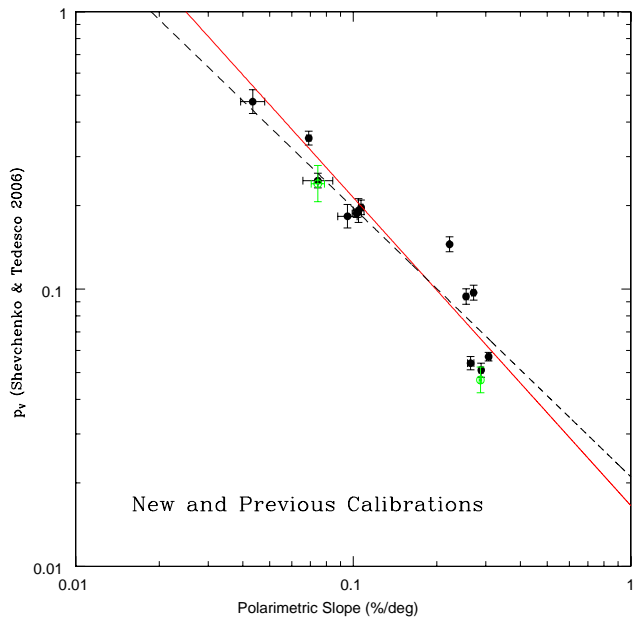}
\end{center}
 \caption{A comparison between the calibration of the slope-albedo relation presented in this paper 
 (red line) and those by \citet{Cellinoetal2012} (black, dashed line).}
 \label{Fig:lupcelfit}
\end{figure}
This seems to be confirmed by Figure \ref{Fig:albalb}, where the albedo
values of the objects considered in our analysis, as they can be
derived from our updated determination of the calibration
coefficients, are plotted \emph{versus} the corresponding albedo
values determined by \citet{ShevTed}. In Fig.$\:$\ref{Fig:diffalb} we
show for each object the difference between the albedo value obtained
from the polarimetric slope and the albedo value given by
\citet{ShevTed}. From Figs.$\:$\ref{Fig:slopealbloglogfit} to 
\ref{Fig:diffalb}, it can be seen that the discrepancies are generally 
low in absolute terms, being mostly below $\pm 0.04$, as shown in
Fig.$\:$\ref{Fig:diffalb}. The error bars of the obtained albedo
values tend to increase with albedo, but this should be expected,
because the slope-albedo relation (Eq.$\:$\ref{Eqn:hpv}) implies that
the error of $p_V$ must increase linearly with $p_V$
itself.\footnote{If we call $y = \log(p_V)$ and $x = \log(h)$, by
  solving by a linear least-squares technique Eq.$\:$\ref{Eqn:hpv}
  and determining the corresponding error $dy$ of $y$, it is easy to
  see that the corresponding error $dp_V$ turns out to be given by
  $dp_V = \ln(10.0) \:p_V \:dy$}

One object turns out to have a polarimetrically-derived albedo that is
significantly discrepant, namely (2) Pallas, (the point which is
located at the highest vertical distance above the best-fit line in
Fig.$~$\ref{Fig:slopealbloglogfit}).  The \citet{ShevTed} albedo value
of this asteroid, $0.145$, seems to be noticeably high for an object
belonging to the $B$ taxonomic class, which is generally supposed to
include low-albedo asteroids. On the other hand, the \citet{ShevTed}
albedo of Pallas is also substantially confirmed by more recent
results based on WISE data \citep{MasieroWISE}. The albedo value
derived from the polarimetric slope turns out to be $0.088 \pm 0.007$,
which would appear to be more in agreement with expectations for a
body belonging to the low-albedo, $B$ class. It is interesting to note
that \citet{DeLeonetal12} observed a large sample of $B$-class objects
and found a continuous variation of near-IR spectral slopes, possibly
suggesting a variety of different compositions. This might 
be a consequence of the fact that the modern $B$ class includes also
some asteroids \citep[the old $F$ class of Tholen, see][]{TholenBar} 
that in the 80s were kept separate based on their behaviour at the 
shortest wavelengths, which are no longer covered in the most modern 
spectroscopic investigations. 
One should also take into account that the surface properties of the largest 
asteroids like (2) Pallas, which retain a larger fraction of the material 
excavated in most impacts, can be different from those of smaller 
asteroids which lose a much larger fraction of the impact debris from most 
collisions to space. We already know, based on their different 
IR beaming parameters, that the surfaces of large and smaller asteroids 
can be significantly different. In any case,
polarimetry seems to indicate for (2) Pallas a low albedo.  In the
absence of any new, updated albedo and/or absolute magnitude value for
this object coming from other sources, we accept this discrepancy. 

Another important problem which is most apparent in
Fig.$\:$\ref{Fig:albalb} concerns the asteroid (64) Angelina, namely
the object having the highest albedo value in our sample. The value
given by \citet{ShevTed} for this object is $0.474 \pm 0.047$. 
According to our new calibration, the resulting albedo turns
out to be $0.540 \pm 0.085$, marginally in agreement with the
\citet{ShevTed} value. The problem here is not this discrepancy
\emph{per se}, but rather the fact that, as shown in
Fig.$\:$\ref{Fig:slopealbloglogfit}, in the high-albedo domain our new
calibration tends to assign albedo values increasingly close to $1.0$
to asteroids having increasingly shallower polarimetric slopes.

In Fig. \ref{Fig:lupcelfit} we show a
comparison between our new slope-albedo relation (shown in red in the
plot) and the most recent previous calibration, namely that
of \citet{Cellinoetal2012} (black line).  Though not being visually very 
different, an effect of the adopted log-log scale, the new calibration
tends, at high values of albedo, to stay well above the values
predicted by the previous calibration. We will show
in a separate paper that in the case of (44) Nysa, an asteroid belonging
to the high-albedo $E$ class, our new calibration assigns to this asteroid
an albedo of about $0.9$, which seems exceedingly high to be credible
for a rocky body. In the domain of low-albedo
asteroids, the larger steepness of the new best-fit line produces
only a marginally better fit with respect to the previous calibration.

This problem of an exceedingly high slope of our linear calibration
seems to be connected with another major problem that is apparent by looking at 
Fig.$\:$\ref{Fig:slopealbloglogfit}. This is the fact that at low albedo,
the data look rather noisy, with some data points, having polarimetric slopes 
between $0.2$\%/deg and $0.3$ \%/deg, which are located well above or below
the linear best-fit of the whole dataset. Trying to fit all the data, 
one is led to accept a linear best-fit whose steepness
is a consequence of the presence of the lowest-albedo asteroids.
This is a problem that has been known since a long time, and we will discuss it 
more extensively in the following Subsection.

\begin{figure}
\begin{center}
\includegraphics[width=88mm]{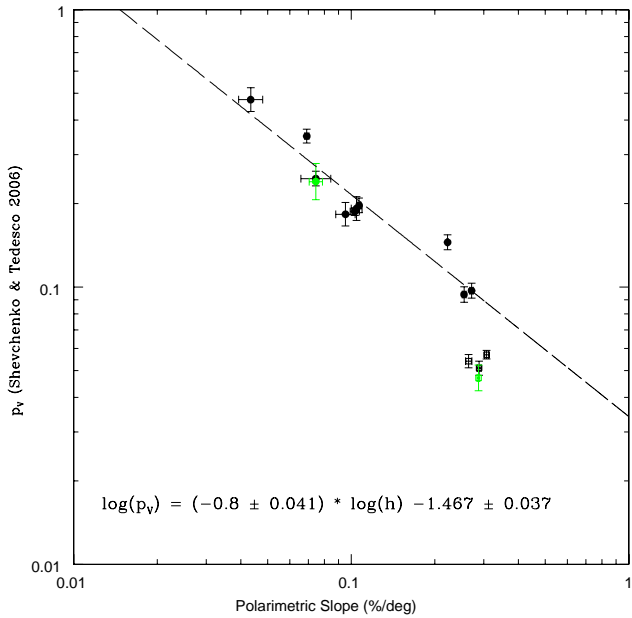}
\end{center}
 \caption{The same as Fig.$~$\ref{Fig:slopealbloglogfit}, but here the 
objects having albedo smaller than $0.08$ (displayed here using open squares) were
not used to derive the displayed linear best-fit. The
corresponding values of the $C_1$ and $C_2$ calibration coefficients are also indicated.}
\label{Fig:slopealbbrightloglogfit}
\end{figure}
\subsection{Effects of excluding low-albedo objects} \label{lowpvslope}
In general terms, looking at Fig.$\:$\ref{Fig:slopealbloglogfit}, one
could be tempted to conclude that it is hard to fit the whole dataset
by using one unique linear relation. In particular, it may
appear that, if one could drop a handful of objects having albedo
lower than about $0.08$, one could obtain a much better fit 
for the remaining objects. This is an old-debated subject, namely 
whether there is evidence of a saturation of the slope -albedo law 
at small albedo values. This
kind of possible saturation is also much more evident in $P_{\rm min}$
- albedo data, as we will see in Section \ref{bestfit}.  

\begin{figure}
\begin{center}
\includegraphics[width=88mm]{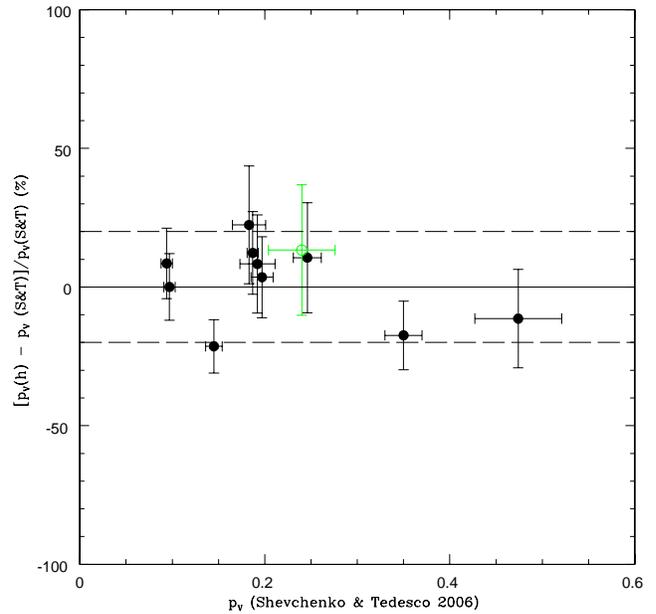}
\end{center}
 \caption{The same as Fig.$~$\ref{Fig:diffalb}, but here the 
objects having albedo smaller than $0.08$ are not included in the analysis, and
the plot shows the relative error of the albedo determinations.}
\label{Fig:diffalbbrightrel}
\end{figure}

What happens if we exlude from our analysis low-albedo
asteroids? Figure \ref{Fig:slopealbbrightloglogfit} shows the results of this
exercise. As expected, the RMS deviation of the linear best-fit of the
data, which is now much shallower than in the case in which we kept all
the available measurements, is quite better, as shown in Table 4.
One could conclude that excluding the asteroids having albedo lower
than $0.08$ is the best way to proceed to obtain an improved, and
quite better, calibration of the slope - albedo law. 
Figure \ref{Fig:diffalbbrightrel} shows that in this way, the relative
error on the derived albedo values turns out to be generally better than
$20$\%, a quite good result.
\begin{figure}
\begin{center}
\includegraphics[width=88mm]{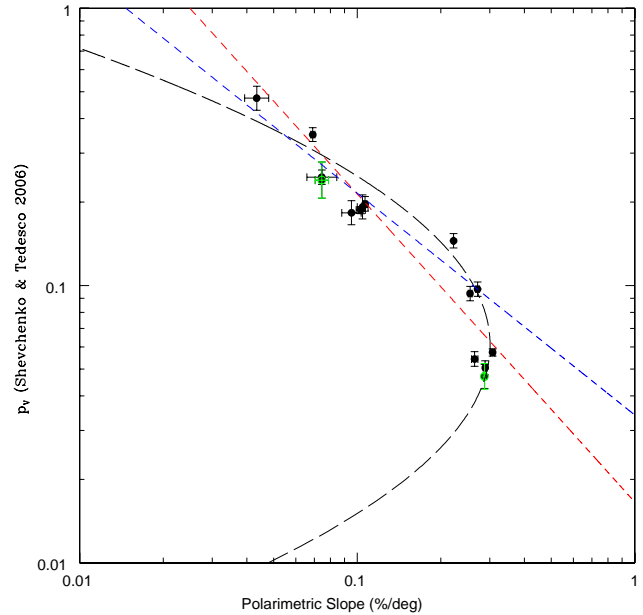}
\end{center}
 \caption{Results of a parabolic best-fit of all available
slope - albedo data. The corresponding linear fits shown in
Fig.$~$\ref{Fig:slopealbloglogfit} and in Fig.$~$\ref{Fig:slopealbbrightloglogfit} 
are also displayed for a comparison (red and blue dashed lines, respectively).}
\label{Fig:Karri}
\end{figure}
The problem, of course,
is that in the practical applications of asteroid polarimetry,
one has at disposal the polarimetric slope derived from observations, 
and wants to derive from it the albedo, which is unknown. Unfortunately, there is
a range of values of polarization slope which is shared by objects having albedo
either around $0.05$ or around $0.10$. By using 
a slope - albedo relation which is not valid for low-albedo asteroids 
may produce a systematic overestimate of the albedo for dark objects.  

The best procedure to be adopted in practice may be the following:
when the polarimetric slope of an object is measured with good
accuracy, it will be better to use the calibration coefficients obtained
by dropping low-albedo objects,
displayed in Table 1 and Fig.$~$\ref{Fig:slopealbbrightloglogfit}, {\em but only
when the polarimetric slope turns out to be smaller than about $0.25$\%/deg}.
In this way, the relative error in the determination of the albedo should be 
within $20$\%, a nice result, as shown in Fig.$~$\ref{Fig:diffalbbrightrel}.

If the slope is larger than the above value, and/or when the value of
polarization slope is more uncertain, the best choice would be 
probably to use the calibration coefficients fitting the whole
population, displayed in Table 1 and Fig.$~$\ref{Fig:slopealbloglogfit}. The
errors that one should expect for the higher values of polarimetric
slopes should be in any case limited, of the order of about $\pm 0.03$,
not negligible in relative terms, but in any case sufficient to 
correctly classify the objects as low-albedo asteroids.

The polarimetric slope data shown in Fig.$~$\ref{Fig:slopealbloglogfit}, can also 
suggest that a linear fit is not fully adequate to represent 
the whole dataset, and a parabolic fit could be more suited to 
better represent the data. 
This is shown in Fig.$~$\ref{Fig:Karri}, in which a parabolic relation
\begin{displaymath}
\log(h) = H_1 (\log(p_V))^2 + H_2 \log(p_V) + H_3
\end{displaymath}
is adopted, and the result of a best-fit procedure
is shown. The linear plots already shown in
Fig.$~$\ref{Fig:slopealbloglogfit} and 
Fig.$~$\ref{Fig:slopealbbrightloglogfit} are also shown for
a visual comparison. It is clear from the Figure 
that a parabolic fit applied to the whole dataset gives good results. 
The slightly worse fit of some objects having albedo around $0.2$
(located below the parabolic fit) seems to be compensated by a much better
fit of the polarimetric slope and albedo of the asteroid (2)
Pallas, which is the most discrepant object when trying linear fits 
to represent the data. However, there is still 
the problem of the ambiguity affecting the values of albedo to be 
assigned to objects having polarimetric slopes between $0.2$\%/deg
and $0.3$\%/deg, which cannot be solved.

For this reason, we list the obtained values of the $A$, $B$, $C$
coefficients of our parabolic fit in Table 1, but we are not claiming
that there is such a big improvement with respect to the classical 
linear fit, specially when dropping the lowest-albedo
objects, to force us to necessarily use a parabolic fit in the future. 
Things could change in the case that the discovery of other objects sharing 
the location of (2) Pallas in the $h$ - $p_V$ plane would confirm that a 
linear fit is not really suited to adequately represent them, even by 
excluding from the analysis low-albedo objects. Another possibility is 
that future theoretical advances
in the interpretation of light scattering phenomena could suggest
that a parabolic fit is intrinsically more correct to fit the
relation between polarimetric slope and geometric albedo, based on 
some physical arguments. 

\section{More on classical approaches: fitting phase-polarization curves}\label{bestfit}

One can wonder whether the slope - albedo relation is really the best
available choice to derive good estimates of asteroid albedos. In
fact, asteroid phase - polarization curves do not include only the
(mostly) linear variation of \pr\ around the inversion angle. A
negative polarization branch also exists, not to mention the behavior
exhibited at large phase angles (not achievable for main belt objects)
by near-Earth asteroids.

Historically, another relation between albedo and polarization
properties was found to involve the $P_{\rm min}$ parameter, as shown
in Eq.$~$(\ref{Eqn:pminpv}). As mentioned in the previous Section, this
has been generally abandoned in recent years, because $P_{\rm min}$
data have been found by some authors to be quite scattered around the
best-fit representation given by Eq.$\:$\ref{Eqn:pminpv}. On the
other hand, one could also wonder whether this might be at least
partly a consequence of the difficulty of deriving accurate values of
$P_{\rm min}$ from the observations, making use of visual
extrapolations of rather sparse polarimetric data.
This leads us to face the problem of finding suitable analytical
representations of the morphology of phase - polarization curves.

\begin{figure}
\begin{center}
\includegraphics[width=88mm]{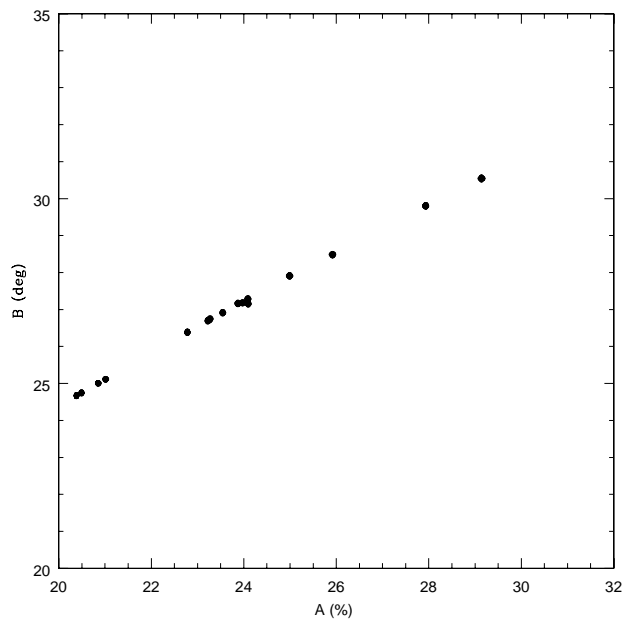}
\end{center}
\caption{\label{Fig:ABkleo} The values of the best-fit parameters $A$ and $B$
found for the best 200 solutions of the linear-exponential relation
applied to polarimetric data of asteroid (216) Kleopatra, using a
``genetic'' algorithm. Note that all the plotted solutions for $A$ and
$B$ give nearly identical residuals and produce best-fit curves
characterized by values of $h$, $P_{\rm min}$ and other polarimetric
parameters which are essentially identical (within the nominal error
bars).}
 \end{figure}
\begin{figure}
\begin{center}
\includegraphics[width=88mm]{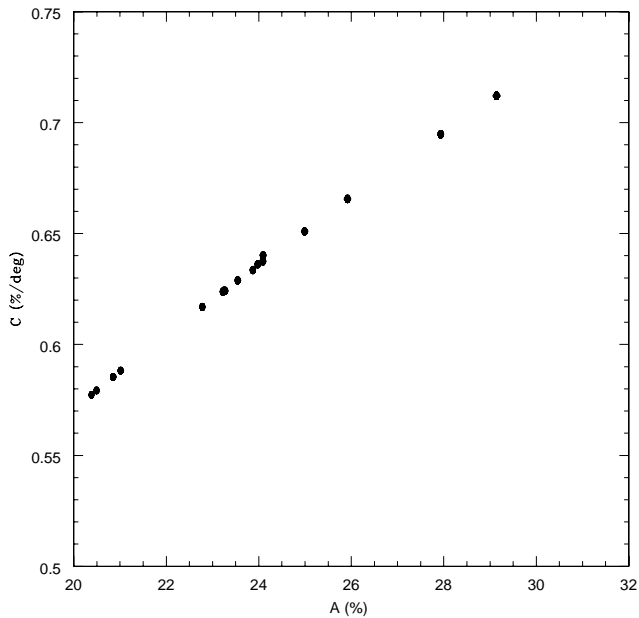}
\end{center}
 \caption{The same as in Fig. \ref{Fig:ABkleo}, but for the $A$ and $C$ parameters in 
Eq. \ref{Eqn:ABC}.}
 \label{Fig:ACkleo}
\end{figure}
In this paper we follow the example of previous authors, and we use the following
exponential-linear relation:
\begin{equation}
\pr = A (e^{-\alpha/B} - 1) + C \cdot \alpha \label{Eqn:ABC}
\end{equation}
where $\alpha$ is the phase angle expressed in degrees, and $A$, $B$,
$C$ are parameters to be determined by means of best-fit
techniques. This analytical representation has been found in the past
to be suited to fit both phase - magnitude relations in asteroid photometry, and 
phase - polarization curves in asteroid polarimetry \citep{Shennaetal03,Muinonenetal09}.
Some examples of practical applications of the above
relation are the best-fit curves of the phase-polarization curves of
the asteroids shown in Fig.$~$\ref{Fig:fig1}.  According to its
mathematical representation, when the parameters $A, B, C$ are all
positive, the exponential-linear relation describes a curve
characterized by a negative polarization branch between $0$ and an
inversion angle $\alpha_{\rm inv}$. The trend tends to become essentially
linear at large phase angles, where the exponential term tends quickly
to zero.

The computation of the best-fit representation of any
phase-polarization curve using Eq.$\:$\ref{Eqn:ABC} can be done in
many ways. In the present paper, we use a genetic algorithm, which, 
starting from a random set of $A,B,C$ values, explores the space of 
possible solution parameters and finds the set of $A,B,C$ values 
producing the smallest possible residuals. Due to the intrinsic properties 
of a genetic approach, the algorithm is launched several times, in order
to have a correspondingly high number of solutions, in order to ensure 
that we are not missing the best possible one.

It should be noted, however, that the evaluation of the errors of
any polarimetric parameter derived by a best-fit of the phase - polarization
curve using Eq.$~$\ref{Eqn:ABC},
is complicated by the fact that the parameters $A$, $B$ and $C$ that
minimise the $\chi^2$ are correlated. To illustrate this situation,
Fig.$~$\ref{Fig:ABkleo} shows that different pairs of $A$ and $B$ values
produce fits nearly indistinguishable in terms of r.m.s. residuals
(the differences being not larger than 0.0015). Figure \ref{Fig:ACkleo} shows
a similar situation for the parameters $A$ and $C$. This means that
the non diagonal elements of the error matrix 
\citep[see, e.g.,][]{Bevington} are not negligible with respect to the 
diagonal ones, and therefore the calculation of the error on any 
polarimetric parameter derived from the exponential-linear curve 
should take into account all the various co-variances. Our method for
the $\chi^2$ minimization is based on a genetic algorithm, which does
not produce automatically the error matrix. For the estimate of the
error on polarimetric parameters derived by best-fit values of 
$A,B,C$, it is therefore more practical to adopt an
alternative method. Let us assume, as an example, that we are interested in determining 
$P_{\rm min}$ and its corresponding uncertainty. The resulting value of 
$P_{\rm min}$ will be the one obtained using the $A,B,C$ values giving the
smallest $\chi^2$. As for the error to be assigned to this determination 
of $P_{\rm min}$, the method that we adopt consists of calculating all $P_{\rm min}$
values corresponding to identified sets of $A,B,C$ parameters that produce a
fit of the phase - polarization curve giving $\chi^2$ values such that 
$\chi^2 \le \chi^2_{\rm min} + 1$. We then define as error on $P_{\rm min}$ 
the half difference between the extremes of the various $P_{\rm min}$ values so
obtained. This approach was followed e.g. by \citet{Bagnuloetal1995}
and is consistent with the error analysis presented by \citet{Bevington}.
Needless to say, this procedure may be applied to the determination of any 
polarimetric parameter (other than $P_{\rm min}$) derived by an 
exponential-linear fit of the phase-polarization curve.

In what follows, since we want to limit our analysis only to high-quality
and well-covered phase - polarization curves, we impose some strict
constraints on the selection of the objects for which we compute a
best-fit using the exponential-linear relation. In particular:
\begin{itemize}

\item we exclude \emph{a priori} from our analysis all measurements
  having a nominal accuracy of \pr\ worse than $0.20$.
\item we require to have at least $4$ accepted measurements taken at
  phase angles $> 2^\circ$.
\item we require to have at least $1$ accepted measurement taken at
  phase angles $\geq 17^\circ$.
\item we require to have at least $1$ accepted measurement taken at
  phase angles $< 14^\circ$.
\item we require to have at least $3$ accepted measurements taken at
  phase angles $< 30^\circ$.
\end{itemize}

As in the case of the computation of the polarimetric slope described
in the previous Section, we limit our analysis to available data
obtained in $V$ filter. Here, however, we add an additional
constraint: we \emph{do not use} for calibration purposes the best-fit
phase - polarization curves of asteroids for which we have fewer than
$10$ accepted \pr\ measurements. All the criteria described above 
are dictated by our will to restrict our
calibration procedures to the objects having excellently determined 
and optimally sampled phase - polarization curves, only.

\subsection{Use of $P_{\rm min}$} \label{Pmin}
Having at disposal a best-fit representation of a phase-polarization
curve according to Eq.$\:$\ref{Eqn:ABC}, one can compute the
resulting $P_{\rm min}$ value and the corresponding phase angle
$\alpha(P_{\rm min})$ at which it is found.  More in detail,
$\alpha(P_{\rm min})$ can be computed by equaling to zero the first
derivative of Eq.$\:$\ref{Eqn:ABC}, from which we obtain:
\begin{displaymath}
\alpha(P_{\rm min}) = \ln(\frac{A}{BC})
\end{displaymath}
then, $P_{\rm min}$ can be computed as $\pr(\alpha(P_{\rm min}))$
using Eq.$\:$\ref{Eqn:ABC}. The nominal uncertainty in $P_{\rm min}$
has to be computed by doing a formal propagation of the errors of the
$A, B, C$ parameters found in the best-fit solution of
xEq.$~$\ref{Eqn:ABC}, using the procedure explained above.

\begin{figure}
\begin{center}
 \vspace{0.5cm}
\includegraphics[width=88mm]{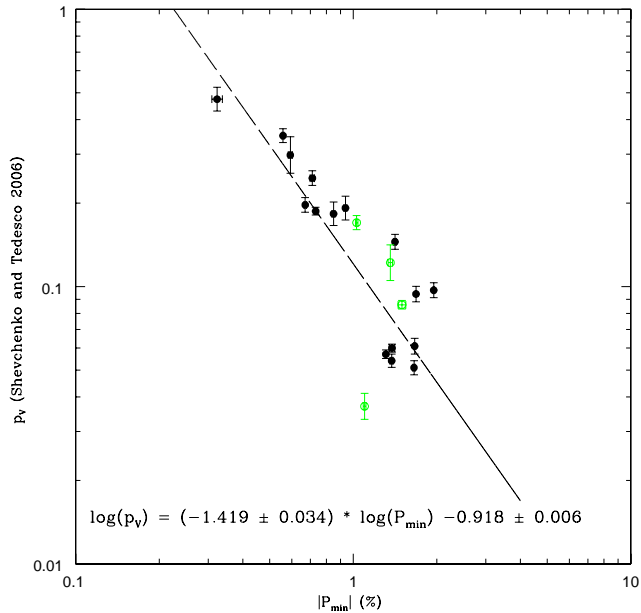}
\end{center}
 \caption{\label{Fig:pminalbloglogfit}
Best-fit relation between $P_{\rm min}$ and 
geometric albedo for an available sample of $20$ asteroid
targets included in the \citet{ShevTed} list.  Only $16$ objects for
which we have at disposal at least ten polarimetric measurements have
been used to derive the best-fit solution. These objects are plotted
as full, black symbols. Open, green symbols represent objects for
which we have fewer than ten observations, and were not used in the 
least-squares computation.
}
\end{figure}

In the case of the $P_{\rm min}$ - albedo relation described by
Eq.$\:$\ref{Eqn:pminpv}, the results of our exercise are shown in
Fig.$\:$\ref{Fig:pminalbloglogfit}, including the best-fit values we
find for the $C_3$ and $C_4$ coefficients.  The calibration is based
on data of the $P_{\rm min}$ values of $16$ objects belonging to the 
\citet{ShevTed} list, for
which we have at least $10$ polarimetric measurements. Four additional
asteroids, for which we have fewer than ten measurements, are also
shown using different symbols, but they were not used in the
computation of the best-fit.  The obtained values of $C_3$ and $C_4$
calibration coefficients, together with their corresponding errors,
are also given in Table 1.
We see that, not unexpectedly, the distribution of the points in the
$P_{\rm min}$ - albedo plane, in our log-log plot (in
Fig.\ref{Fig:pminalbloglogfit}) makes it difficult to find a
satisfactory linear fit.  Correspondingly, the agreement of the
resulting albedos with those of the \citet{ShevTed} list is 
significantly worse than in the case of the calibration based on the 
polarimetric slope. As shown in Table \ref{Tab:Tab3}, we find a large 
discrepancy in the case of the bright asteroid (64) Angelina, for which a 
very high value of albedo of $0.600 \pm 0.044$
is obtained from the $P_{\rm min}$ - albedo relation, much larger than
the $0.474 \pm 0.047$ value listed by \citet{ShevTed}. The majority of
the other asteroids of our sample, conversely, tends to have an albedo
underestimated with respect to the calibration values, apart from a few ones 
having the lowest albedo values. Among them, we note
the difference between the very low albedo value found by
\citet{ShevTed} for asteroid (444) Gyptis ($0.037 \pm 0.004$) and the
value of $0.106 \pm 0.002$ that we obtain from its $P_{\rm min}$ value.
This object, however, was not used in the derivation of the best-fit
computation, because only $6$ polarimetric measurements are currently 
available for it, and moreover the phase - polarization curve (not shown)
is quite noisy. We note also that the albedo value found by \citet{ShevTed} 
is very low, and might possibly be slightly underestimated.

\begin{figure}
\begin{center}
 \vspace{0.5cm}
\includegraphics[width=88mm]{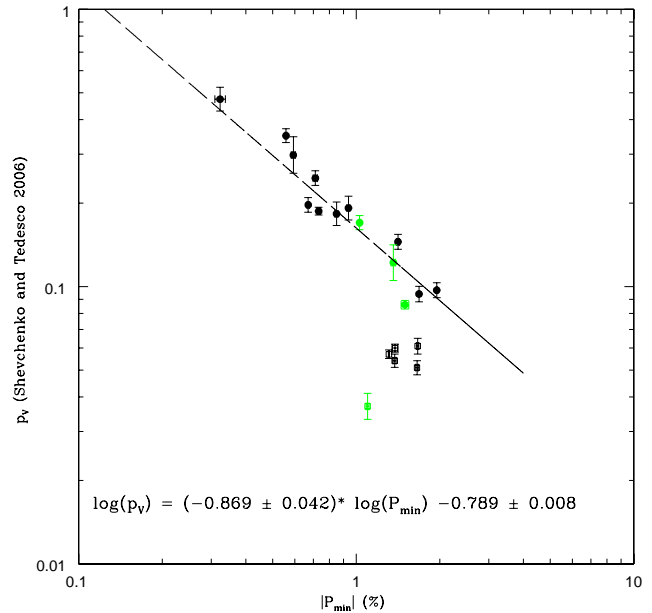}
\end{center}
 \caption{\label{Fig:pminalbbrightloglogfit}
The same as Fig.$~$\ref{Fig:pminalbloglogfit}, but here the objects
having albedo lower than $0.08$ (sowed using open symbols) were not
used in the computation of the linear best-fit.}
\end{figure}
By simply looking at Fig.\ref{Fig:pminalbloglogfit},
a saturation of the $P_{\rm min}$ - albedo relation at low albedo values 
is even more evident than in the case of the slope - albedo relation 
analyzed in Section$~$\ref{lowpvslope}. It seems therefore that a removal of asteroids
having albedo less than about $0.08$ from the best-fit computation is 
even more justified in this case. The result of this exercise is shown in 
Fig.$~$\ref{Fig:pminalbbrightloglogfit}. The improvement of the
residuals, listed in Table 4, is very important, as also visually shown 
in the Figure. 

The big improvement of the obtained best-fit makes this $P_{\rm min}$ - albedo 
relation much more suitable for the determination of the albedo, but, again,
this refers to only a more limited interval of possible
$P_{\rm min}$ values, in particular those lower (in absolute value) than about $1$\%. 
For objects having deeper $P_{\rm min}$, the corresponding interval of  
possible albedo values is
exceedingly wide to be used to derive a useful albedo determination.
Based on our results, we confirm therefore that, in general, the use of 
$P_{\rm min}$ as a reliable diagnostic of the albedo, 
but for asteroids exhibiting a shallow polarization branch,
should not be encouraged.

\subsection{An alternative derivation of the polarimetric slope} \label{hABC}
Using the global fitting of the phase polarization curves given by
Eq.$\:$\ref{Eqn:ABC}, it is also possible to modify the way to derive
the polarimetric slope $h$. In so doing, one can make use of all the
available polarimetric measurements, and not only of those obtained in
a more or less narrow interval of phase angles centered around the
inversion angle. In particular, the polarimetric slope can be
determined as the first derivative of \pr\ with respect to
the phase angle (using Eq.$\:$\ref{Eqn:ABC}), where the derivative
has to be computed at the inversion angle $\alpha_{\rm inv}$. In turn, the
value of $\alpha_{\rm inv}$ can also be derived with excellent accuracy
from Eq.$\:$\ref{Eqn:ABC}. We adopt here a very simple numerical
approach. Having determined the values of parameters $A$, $B$, $C$
corresponding to the nominal best-fit solution, we make an iterative
computation of \pr\, starting from an initial phase angle value of
$1$ degree, and using a fixed increment of $+0.02$ deg in phase. When
$\pr(i+1) \cdot \pr(i)$ becomes negative, we consider that
$\alpha_{inv}$ is equal to $\pr(i) + 0.01$ degrees, with an
uncertainty of $0.02$ degrees.

\begin{figure}
\begin{center}
\includegraphics[width=88mm]{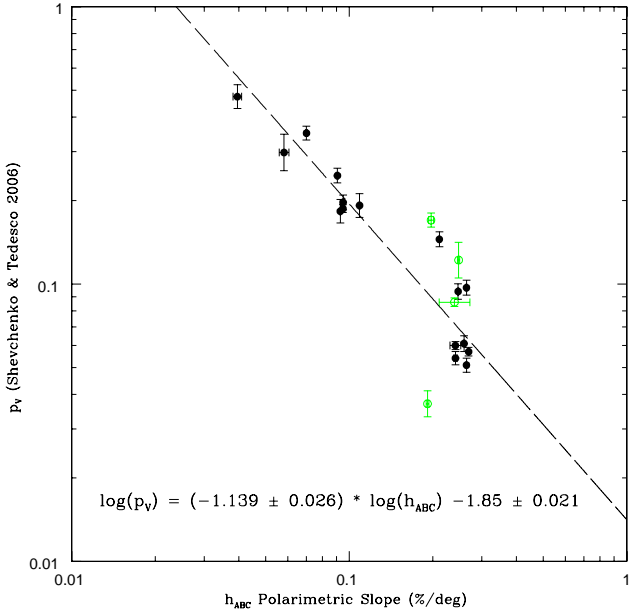}
\end{center}
 \caption{The same as Fig.$\:$\ref{Fig:slopealbloglogfit}, but here the polarimetric slope
 $h_{ABC}$ is computed as the first derivative of \pr\ (according to Eq.$\:$\ref{Eqn:ABC}) at 
 the inversion angle $\alpha_{inv}$ for $20$ asteroids of the \citet{ShevTed} list. Four 
 objects having fewer than ten polarimetric measurements were not used
 for the computation of the best-fit. They are indicated by open, green symbols.}
 \label{Fig:slopeABCalbloglogfit}
\end{figure}
\begin{figure}
\begin{center}
\includegraphics[width=88mm]{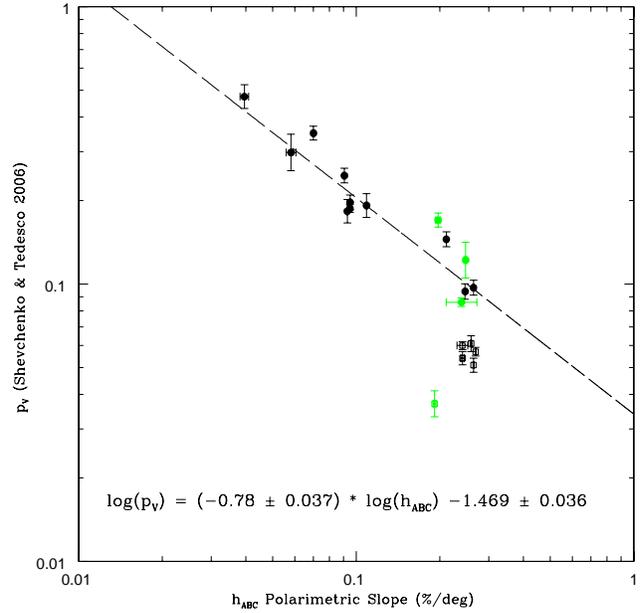}
\end{center}
 \caption{The same as Fig.$\:$\ref{Fig:slopeABCalbloglogfit}, but here 
objects having an albedo lower than $0.08$ (shown using open squares) are 
not used in the computation of the best-fit.}
 \label{Fig:slopeABCalbbrightloglogfit}
\end{figure}
Having determined the inversion angle, we can compute the resulting
polarimetric slope as the first derivative of Eq.$\:$\ref{Eqn:ABC},
using the same procedure already adopted for $P_{\rm min}$ to derive
its nominal uncertainty. In what follows we will always call $h_{ABC}$ 
these new value of the slope computed as explained above. We
obtained $h_{ABC}$ values for $20$ objects of the \citet{ShevTed}
list, as shown in Fig.\ref{Fig:slopeABCalbloglogfit}. Also in this
case, however, we did not use in the best-fit computation the data
of four objects having fewer than $10$ polarimetric
measurements. These four asteroids, namely (78) Diana, (216)
Kleopatra, (431) Nephele and (444) Gyptis, are indicated in
Fig.\ref{Fig:slopeABCalbloglogfit} by means of open, green symbols.
Note that our sample includes now three extra objects, (27) Euterpe,
(41) Daphne and (47) Aglaja, which did not satisfy our previous
acceptability criterion for the computation of the polarimetric slope
$h$ carried out in Section \ref{classical}. Conversely, in Section
\ref{classical} we made use of polarimetric slopes for the two
asteroids (105) Artemis and (124) Alkeste, which do not satisfy our
criteria for the computation of $h_{ABC}$.

The resulting values of the calibration coefficients $C_1$ and $C_2$
are shown in Fig.$~$\ref{Fig:slopeABCalbloglogfit}, and they are also
listed, together with their errors, in Table 1.
As in the cases seen above, we also computed the best-fit values of the 
calibration parameters which are obtained by removing from the computation
the asteroids having albedo lower than $0.08$. The results of this 
exercise, listed in Table 1, are also shown in 
Fig.$~$\ref{Fig:slopeABCalbbrightloglogfit}.

By looking at the results, we find that the linear best-fit of
the slope - albedo relation look, again, reasonably good. However, there is 
not any improvement with respect to the case when the polarimetric slope was
computed by doing a more trivial linear fit of the available data
around the inversion angle (see Fig.$\:$\ref{Fig:albalb}).  As
opposite, the RMS deviation with respect to the \citet{ShevTed} albedos turn out 
to be slightly worse, as shown in Table 4.

\begin{table*} 
\label{Tab:Tab1}
	\centering
	\caption{Resulting values and nominal uncertainties of the calibration coefficients in 
	different albedo - polarization relations considered in this paper. The result of a possible
        parabolic fit of the polarimetric slope as a function of albedo is also listed (see text).}
	\small
		\begin{tabular}{lrrr}
		& & \\
  $\log(p_V) = C_1 \log(h) + C_2$                          & $ C_1 = -1.111 \pm 0.031$ & $C_2 = -1.781 \pm 0.025$ \\ 
  $\log(p_V) = C_1 \log(h) + C_2$ ($p_V \ge 0.08$)         & $ C_1 = -0.800 \pm 0.041$ & $C_2 = -1.467 \pm 0.037$ \\ 
  $\log(p_V) = C_1 \log(h_{ABC}) + C_2$                     & $ C_1 = -1.139 \pm 0.026$ & $C_2 = -1.850 \pm 0.021$ \\
  $\log(p_V) = C_1 \log(h_{ABC}) + C_2$ ($p_V \ge 0.08$)    & $ C_1 = -0.780 \pm 0.037$ & $C_2 = -1.469 \pm 0.036$ \\  
  $\log(p_V) = C_3 \log(P_{\rm min}) + C_4$                  & $ C_3 = -1.419 \pm 0.034$ & $C_4 = -0.918 \pm 0.006$ \\
  $\log(p_V) = C_3 \log(P_{\rm min}) + C_4$ ($p_V \ge 0.08$) & $ C_3 = -0.869 \pm 0.042$ & $C_4 = -0.789 \pm 0.008$ \\
  $\log(p_V) = C_{\psi1} \log(\Psi) + C_{\psi2}$               & $ C_{\psi1} = -0.987 \pm 0.022$ & $C_{\psi2} = -0.458 \pm 0.013$ \\ 
  $\log(p_V) = C\ast_1\: p\!\ast +\: C\ast_2$              & $ C\ast_1 = -0.896 \pm 0.029$ & $C\ast_2 = -1.457 \pm 0.018$ \\
  $\log(h) = H_1 (\log(p_V))^2 + H_2 \log(p_V) + H_3$ & $H_1 =-1.294 \pm 0.001$ & $H_2 = -3.140 \pm 0.001$ & 
                                                           $H_3 = -2.428 \pm 0.001$ \\ 
		& &
    \end{tabular}
\end{table*}
\begin{figure}
\begin{center}
\includegraphics[width=88mm]{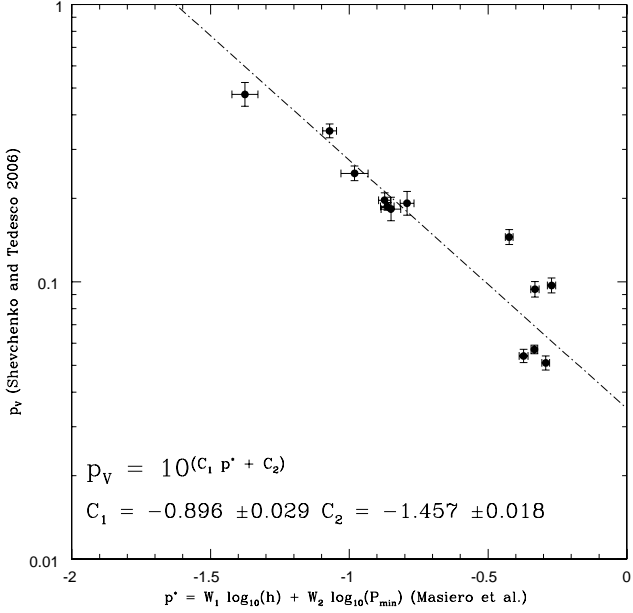}
\end{center}
 \caption{The $p\ast$-albedo relation for the $13$ asteroids of the list of \citet{ShevTed} for which a
 reliable estimate of $p\ast$ is possible based on obtained values of the $h$ polarimetric slope 
 and of $P_{\rm min}$. The best-fit solution is plotted together with the individual data. The newly derived 
 values of the $A$ and $B$ parameters and their resulting errors are also indicated.}
 \label{Fig:pstarfit}
\end{figure}
\begin{figure}
\begin{center}
\includegraphics[width=88mm]{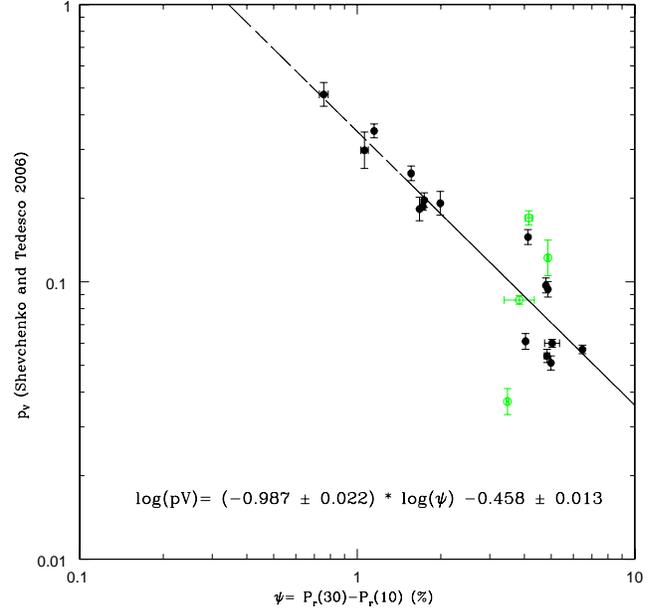}
\end{center}
 \caption{The $\Psi$-albedo relation in log - log scale for $20$ asteroids of the list of \citet{ShevTed} 
for which  we were able to compute the $\Psi$ parameter. Four asteroids having fewer than $10$ polarimetric 
measurements were not used in the computation of the linear best-fit and are indicated by open, green 
symbols. The best-fit solution corresponding to the relation $\log(p_V) = C1 \:\log(\Psi) +C_2$ is plotted 
together with the individual data. The best-fit values of the parameters and their nominal uncertainties 
are also indicated in Table 1.}
 \label{Fig:psialbloglogfit}
\end{figure}
\begin{figure}
\begin{center}
\includegraphics[width=88mm]{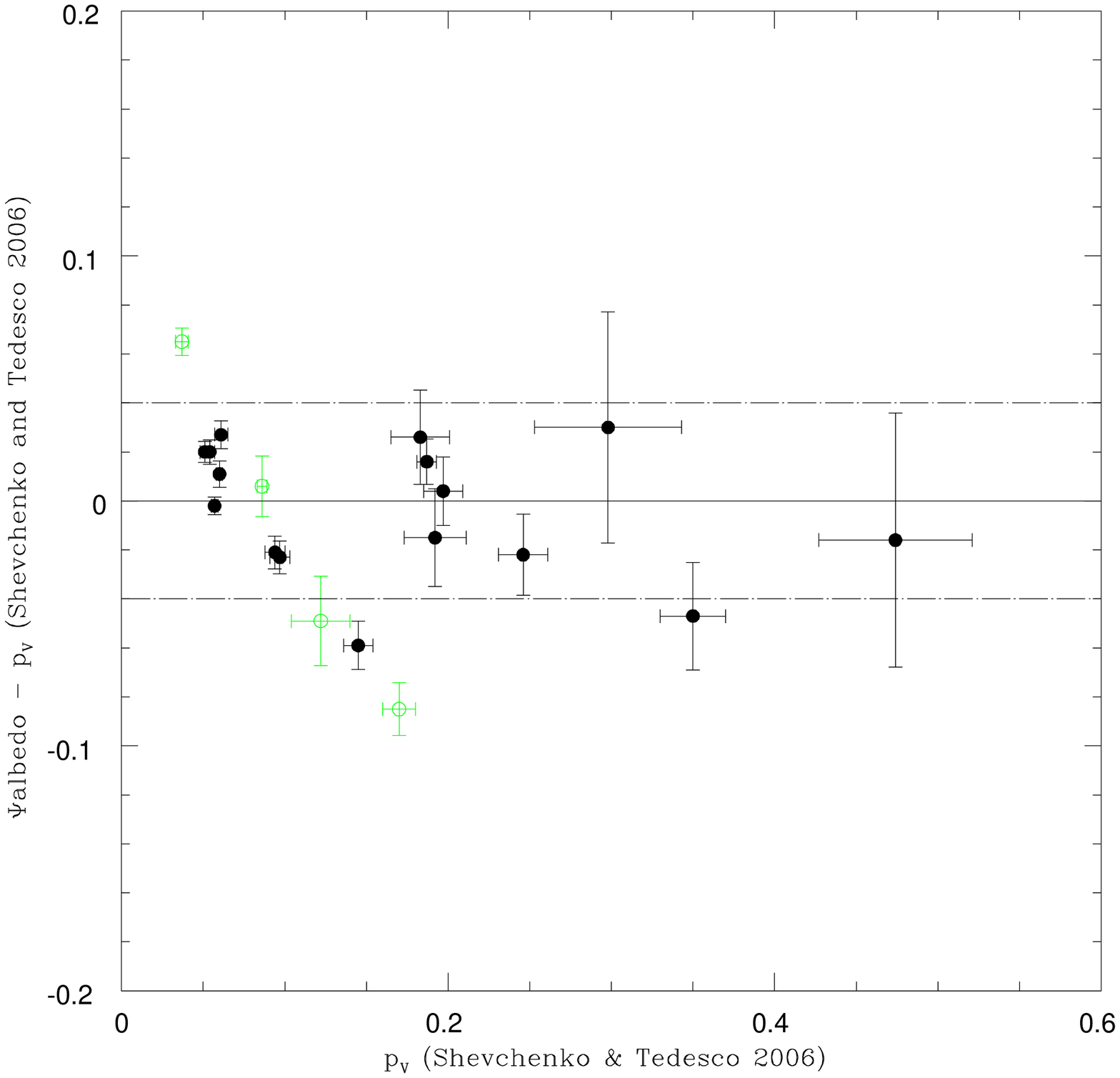}
\end{center}
 \caption{Differences between the albedos of $20$ objects, derived using
 our calibration of the $\Psi$ - albedo relation, and the corresponding albedo 
 values given by \citet{ShevTed}. Four asteroids having fewer than $10$ polarimetric measurements
 were not used in the computation of the linear best-fit and are indicated by open, green symbols.}
 \label{Fig:diffalbpsi}
\end{figure}

The computation of the polarimetric slope from a simple linear fit of data
distributed around the inversion angle, or from the computation of
the first derivative of \pr computed at the inversion angle, 
which would better correspond to the ideal definition of this
parameter, is therefore not fully equivalent. It turns out that, opposite to our
own expectations, the simpler (purely linear) approach seems to give
slightly better results, in spite of all the uncertainties.

The results of the exercises described in this Section are summarized
in Table 2, in which we list all the polarimetric parameters
considered in our analysis (including some which will be explained in
the next Section), and in Table 3, where we list the \citet{ShevTed}
values of albedo, together with the corresponding values of albedo
derived from the considered polarimetric parameters and their nominal
errors. Note that in Table 2, we always give for the inversion angle
$\alpha_{\rm inv}$ the value obtained from the best-fit of the whole phase
- polarization curve using the exponential - linear relation. Only in
cases in which this is not available, but we have at disposal a
polarimetric slope $h$ obtained as described in Section
\ref{classical}, we assign to $\alpha_{\rm inv}$ the value corresponding
to the intersection of the polarimetric slope with the $\pr = 0$ line.

The polarimetric parameters $h$, $h_{ABC}$ and $P_{\rm min}$, and 
corresponding albedos obtained from calibrations based on all and 
only the objects having albedos larger than $0.08$,
as discussed in this and the previous Section, are also listed in 
Table 5. The improvement of the agreement between the albedos
obtained from polarimetric parameters and the albedos given by
\citet{ShevTed} is evident. Our considerations about the best possible use
of these calibrations have been already exposed in previous Sections.  

\section{Other polarimetric parameters}\label{altapp}

The failure of our attempt to obtain more accurate albedo values by
using a polarimetric slope ($h_{ABC}$) obtained by a formal
computation of the first derivative of Eq.$\:$\ref{Eqn:ABC} at the
inversion angle, can be important.  A much simpler linear fit of
polarimetric data spread over a large interval of phase angles, seems
to be capable of giving slightly more accurate albedo solutions.  This
can be an indication that using polarimetric
parameters describing the behavior of the phase - polarization curve
only at some single value of phase angle, or in a limited portion of
the phase angle interval, for the determination of the geometric albedo, 
could be not a very good idea.

\begin{table*} 
\label{Tab:Tab2}
\centering
\caption{Summary of the polarimetric parameters found for all asteroids included in the \citet{ShevTed} 
list, for which we have a suitable coverage of the phase - polarization curves. Each asteroid is 
identified by its number $N$. The second column gives the number $N_{\rm obs}$ of polarimetric measurements 
used in the analysis. For the meaning of the other parameters, see the text.} 
\begin{tabular}{rrccrcccr} & 
      &                   &                &                &                  &                  &                   & \\
\multicolumn{1}{c}{$N$} & \multicolumn{1}{c}{$N_{\rm obs}$} & \multicolumn{1}{c}{$h$} & \multicolumn{1}{c}{$\alpha_{\rm inv}$} & 
\multicolumn{1}{c}{$\alpha(P_{\rm min})$} &
\multicolumn{1}{c}{$P_{\rm min}$} & \multicolumn{1}{c}{$\Psi$} & \multicolumn{1}{c}{$h_{ABC}$} & \multicolumn{1}{c}{$p\!\ast$} \\
      &        &           &                &                &                  &                  &                   & \\
    1 & 33 &  0.2549 $\pm$ 0.0041 & 18.13 $\pm$ 0.02 &  7.20 $\pm$ 0.02 & -1.683 $\pm$ 0.005 & 4.863 $\pm$ 0.020 & 0.2467 $\pm$  
		0.0009 &  0.331 $\pm$ 0.015\\
    2 & 22 &  0.2223 $\pm$ 0.0015 & 18.57 $\pm$ 0.02 &  7.59 $\pm$ 0.07 & -1.413 $\pm$ 0.017 & 4.123 $\pm$ 0.010 & 0.2111 $\pm$ 
		0.0004 & -0.424 $\pm$ 0.014\\
    3 & 26 &  0.1029 $\pm$ 0.0039 & 20.31 $\pm$ 0.02 &  8.04 $\pm$ 0.08 & -0.732 $\pm$ 0.009 & 1.724 $\pm$ 0.023 & 0.0949 $\pm$ 
		0.0013 & -0.863 $\pm$ 0.024\\
    4 & 26 &  0.0691 $\pm$ 0.0012 & 22.37 $\pm$ 0.02 &  9.22 $\pm$ 0.08 & -0.558 $\pm$ 0.005 & 1.150 $\pm$ 0.006 & 0.0701 $\pm$ 
		0.0005 & -1.071 $\pm$ 0.025\\
    8 & 28 &  0.1067 $\pm$ 0.0016 & 20.05 $\pm$ 0.02 &  8.30 $\pm$ 0.03 & -0.671 $\pm$ 0.007 & 1.743 $\pm$ 0.006 & 0.0950 $\pm$ 
		0.0003 & -0.873 $\pm$ 0.021\\
   27 & 12 & \multicolumn{1}{c}{--} & 21.51 $\pm$ 0.02 &  7.27 $\pm$ 0.22 & -0.593 $\pm$ 0.010 & 1.062 $\pm$ 0.033 & 0.0582 $\pm$ 
	  0.0023 & \multicolumn{1}{c}{--} \\
   39 & 20 &  0.0745 $\pm$ 0.0093 & 21.43 $\pm$ 0.02 &  8.66 $\pm$ 0.07 & -0.712 $\pm$ 0.013 & 1.563 $\pm$ 0.017 & 0.0906 $\pm$ 
	  0.0009 & -0.981 $\pm$ 0.049\\
   41 & 11 & \multicolumn{1}{c}{--} & 22.53 $\pm$ 0.02 & 10.54 $\pm$ 0.04 & -1.664 $\pm$ 0.015 & 4.040 $\pm$ 0.034 & 0.2591 $\pm$ 
	  0.0019 & \multicolumn{1}{c}{--}\\ 
   47 & 11 & \multicolumn{1}{c}{--} & 17.95 $\pm$ 0.02 &  7.90 $\pm$ 0.20 & -1.378 $\pm$ 0.031 & 5.037 $\pm$ 0.312 & 0.2412 $\pm$ 
	  0.0109 & \multicolumn{1}{c}{--} \\
   51 & 20 &  0.2713 $\pm$ 0.0040 & 20.35 $\pm$ 0.02 &  8.29 $\pm$ 0.05 & -1.950 $\pm$ 0.015 & 4.784 $\pm$ 0.023 & 0.2644 $\pm$ 
	  0.0013 & -0.271 $\pm$ 0.015\\ 
   64 & 13 &  0.0434 $\pm$ 0.0043 & 18.77 $\pm$ 0.02 &  6.75 $\pm$ 0.26 & -0.323 $\pm$ 0.014 & 0.757 $\pm$ 0.028 & 0.0395 $\pm$ 
	  0.0014 & -1.376 $\pm$ 0.048\\ 
   78 &  5 & \multicolumn{1}{c}{--} & 22.19 $\pm$ 0.02 & 10.42 $\pm$ 0.29 & -1.497 $\pm$ 0.048 & 3.835 $\pm$ 0.481 & 0.2394 $\pm$ 
	  0.0307 & \multicolumn{1}{c}{--} \\
   85 & 11 &  0.2648 $\pm$ 0.0071 & 19.07 $\pm$ 0.02 &  8.68 $\pm$ 0.16 & -1.375 $\pm$ 0.017 & 4.825 $\pm$ 0.108 & 0.2414 $\pm$ 
	  0.0034 & -0.372 $\pm$ 0.016\\ 
  105 &  6 &  0.2872 $\pm$ 0.0032 & 19.75 $\pm$ 0.31 &  \multicolumn{1}{c}{--}& \multicolumn{1}{c}{--} & \multicolumn{1}{c}{--} &
	\multicolumn{1}{c}{--}    & \multicolumn{1}{c}{--} \\ 
  124 &  5 &  0.0745 $\pm$ 0.0041 & 19.79 $\pm$ 1.46 & \multicolumn{1}{c}{--} & \multicolumn{1}{c}{--} & \multicolumn{1}{c}{--} & 
	\multicolumn{1}{c}{--} & \multicolumn{1}{c}{--} \\ 
  129 & 10 &  0.0953 $\pm$ 0.0077 & 21.07 $\pm$ 0.02 &  7.61 $\pm$ 0.03 & -0.849 $\pm$ 0.011 & 1.676 $\pm$ 0.009 & 0.0928 $\pm$ 
	  0.0004 & -0.850 $\pm$ 0.035\\  
  216 &  9 & \multicolumn{1}{c}{--} & 18.83 $\pm$ 0.02 &  8.93 $\pm$ 0.09 & -1.028 $\pm$ 0.011 & 4.151 $\pm$ 0.124 & 0.1971 $\pm$ 
	  0.0041 & \multicolumn{1}{c}{--} \\ 
  230 & 15 &  0.1045 $\pm$ 0.0047 & 20.45 $\pm$ 0.02 &  7.55 $\pm$ 0.06 & -0.937 $\pm$ 0.012 & 1.991 $\pm$ 0.017 & 0.1089 $\pm$ 
	  0.0010 & -0.792 $\pm$ 0.025\\ 
  324 & 20 &  0.2890 $\pm$ 0.0046 & 19.73 $\pm$ 0.02 &  8.70 $\pm$ 0.05 & -1.655 $\pm$ 0.016 & 4.987 $\pm$ 0.024 & 0.2644 $\pm$ 
	  0.0010 & -0.292 $\pm$ 0.014\\ 
  431 &  7 & \multicolumn{1}{c}{--} & 19.67 $\pm$ 0.02 &  9.29 $\pm$ 0.07 & -1.360 $\pm$ 0.024 & 4.864 $\pm$ 0.048 & 0.2478 $\pm$ 
	  0.0023 & \multicolumn{1}{c}{--} \\
  444 &  6 & \multicolumn{1}{c}{--} & 20.65 $\pm$ 0.02 &  9.77 $\pm$ 0.04 & -1.098 $\pm$ 0.012 & 3.477 $\pm$ 0.041 & 0.1914 $\pm$ 
	  0.0021 & \multicolumn{1}{c}{--} \\ 
  704 & 32 &  0.3074 $\pm$ 0.0054 & 15.73 $\pm$ 0.02 &  7.02 $\pm$ 0.01 & -1.310 $\pm$ 0.006 & 6.478 $\pm$ 0.011 & 0.2692 $\pm$ 
	  0.0006 & -0.333 $\pm$ 0.012\\ 
      &         &          &                &                &                  &                  &                   & 
\end{tabular}
\end{table*}	
\begin{table*} 
\label{Tab:Tab3}
\centering
\caption{Resulting albedo values $p_V$ for all asteroids belonging to the \citet{ShevTed} list, for which we have polarimetric 
observations suitable to derive an albedo value using one or more of the techniques explained in the text. Only the 
albedo corresponding to the calibration computed using all available objects (not only those with $p_V > 0.08$) are listed.
The last column gives, for a comparison, the albedo given by \citet{ShevTed}.} 
\begin{tabular}{rcccccc}
 & & & \\
\multicolumn{1}{c}{Number} & $p_V(h)$ & $p_V(h_{ABC})$ & $p_V(P_{\rm min})$ & $p_V(\Psi)$ & $p_V(p\!\ast)$ & $p_V$(S\&T) \\ 
 & & & \\
    1 & 0.076 $\pm$ 0.006 & 0.070 $\pm$ 0.004 & 0.058 $\pm$ 0.001 & 0.073 $\pm$ 0.003 & 0.069 $\pm$ 0.004 & 0.094 $\pm$ 0.006\\
    2 & 0.088 $\pm$ 0.007 & 0.083 $\pm$ 0.005 & 0.074 $\pm$ 0.002 & 0.086 $\pm$ 0.004 & 0.084 $\pm$ 0.005 & 0.145 $\pm$ 0.009\\
    3 & 0.207 $\pm$ 0.021 & 0.206 $\pm$ 0.016 & 0.188 $\pm$ 0.005 & 0.203 $\pm$ 0.007 & 0.207 $\pm$ 0.018 & 0.187 $\pm$ 0.006\\
    4 & 0.322 $\pm$ 0.033 & 0.292 $\pm$ 0.025 & 0.276 $\pm$ 0.008 & 0.303 $\pm$ 0.009 & 0.318 $\pm$ 0.031 & 0.350 $\pm$ 0.020\\
    8 & 0.199 $\pm$ 0.018 & 0.206 $\pm$ 0.016 & 0.213 $\pm$ 0.005 & 0.201 $\pm$ 0.007 & 0.212 $\pm$ 0.018 & 0.197 $\pm$ 0.012\\
   27 & \multicolumn{1}{c}{--} & 0.360 $\pm$ 0.036 & 0.254 $\pm$ 0.008 & 0.328 $\pm$ 0.014 & \multicolumn{1}{c}{--} & 0.298 $\pm$ 
	0.045\\
   39 & 0.297 $\pm$ 0.051 & 0.218 $\pm$ 0.017 & 0.196 $\pm$ 0.006 & 0.224 $\pm$ 0.007 & 0.264 $\pm$ 0.034 & 0.246 $\pm$ 0.015\\
   41 & \multicolumn{1}{c}{--} & 0.066 $\pm$ 0.004 & 0.059 $\pm$ 0.001 & 0.088 $\pm$ 0.004 & \multicolumn{1}{c}{--} & 0.061 $\pm$ 
	0.004\\
   47 & \multicolumn{1}{c}{--} & 0.071 $\pm$ 0.006 & 0.077 $\pm$ 0.003 & 0.071 $\pm$ 0.005 & \multicolumn{1}{c}{--} & 0.060 $\pm$ 
	0.002\\
   51 & 0.071 $\pm$ 0.005 & 0.064 $\pm$ 0.004 & 0.047 $\pm$ 0.001 & 0.074 $\pm$ 0.003 & 0.061 $\pm$ 0.003 & 0.097 $\pm$ 0.006\\
   64 & 0.540 $\pm$ 0.085 & 0.560 $\pm$ 0.059 & 0.600 $\pm$ 0.044 & 0.458 $\pm$ 0.022 & 0.597 $\pm$ 0.084 & 0.474 $\pm$ 0.047\\
   78 & \multicolumn{1}{c}{--} & 0.072 $\pm$ 0.011 & 0.068 $\pm$ 0.003 & 0.092 $\pm$ 0.012 & \multicolumn{1}{c}{--} & 0.086 $\pm$ 
	0.003\\
   85 & 0.072 $\pm$ 0.006 & 0.071 $\pm$ 0.004 & 0.077 $\pm$ 0.002 & 0.074 $\pm$ 0.004 & 0.075 $\pm$ 0.004 & 0.054 $\pm$ 0.003\\
  105 & 0.066 $\pm$ 0.005 & \multicolumn{1}{c}{--} & \multicolumn{1}{c}{--} & \multicolumn{1}{c}{--} & \multicolumn{1}{c}{--} &
	0.047 $\pm$ 0.005 \\
  124 & 0.297 $\pm$ 0.034 & \multicolumn{1}{c}{--} & \multicolumn{1}{c}{--} & \multicolumn{1}{c}{--} & \multicolumn{1}{c}{--} &
	0.240 $\pm$ 0.036 \\
  129 & 0.226 $\pm$ 0.029 & 0.212 $\pm$ 0.017 & 0.152 $\pm$ 0.004 & 0.209 $\pm$ 0.007 & 0.202 $\pm$ 0.020 & 0.183 $\pm$ 0.018\\
  216 & \multicolumn{1}{c}{--} & 0.090 $\pm$ 0.006 & 0.116 $\pm$ 0.002 & 0.085 $\pm$ 0.004 & \multicolumn{1}{c}{--} & 0.170 $\pm$ 
	0.010\\
  230 & 0.204 $\pm$ 0.021 & 0.177 $\pm$ 0.013 & 0.132 $\pm$ 0.003 & 0.177 $\pm$ 0.006 & 0.179 $\pm$ 0.015 & 0.192 $\pm$ 0.019\\
  324 & 0.066 $\pm$ 0.005 & 0.064 $\pm$ 0.004 & 0.059 $\pm$ 0.002 & 0.071 $\pm$ 0.003 & 0.064 $\pm$ 0.003 & 0.051 $\pm$ 0.003\\
  431 & \multicolumn{1}{c}{--} & 0.069 $\pm$ 0.004 & 0.078 $\pm$ 0.002 & 0.073 $\pm$ 0.003 & \multicolumn{1}{c}{--} & 0.122 $\pm$ 
	0.018\\
  444 & \multicolumn{1}{c}{--} & 0.093 $\pm$ 0.006 & 0.106 $\pm$ 0.002 & 0.102 $\pm$ 0.004 & \multicolumn{1}{c}{--} & 0.037 $\pm$ 
	0.004\\
  704 & 0.061 $\pm$ 0.004 & 0.063 $\pm$ 0.004 & 0.082 $\pm$ 0.001 & 0.055 $\pm$ 0.003 & 0.069 $\pm$ 0.004 & 0.057 $\pm$ 0.002\\
 & & &
\end{tabular}
\end{table*}
\begin{table*}
\label{Tab:Tab4}
	\centering
	\caption{Average RMS deviation of polarimetrically-derived albedo with respect to the values 
                 in the \citet{ShevTed} list, using different possible albedo - polarization relations 
                 described in the text. For each case, $N$ is the number of observed asteroids used to 
                 obtain the calibration}
		\begin{tabular}{llcr}
		& & & \\
Albedo computed from: & & \multicolumn{1}{c}{RMS deviation}  & \multicolumn{1}{c}{$N$}\\
		& & & \\
$h$ slope from linear fit & (all objects)                                      & $0.035$ & $15$ \\
$h$ slope from linear fit & (only objects having $p_V > 0.08$)                 & $0.033$ & $11$ \\
$h_{abc}$ slope from linear-exponential fit & (all objects)                     & $0.038$  & $16$ \\
$h_{abc}$ slope from linear-exponential fit & (only objects having $p_V > 0.08$) & $0.034$  & $11$ \\
$P_{\rm min}$ & (all objects)                                                     & $0.051$ & $16$ \\
$P_{\rm min}$ & (only objects having $p_V > 0.08$)                                & $0.035$ & $11$ \\
$\Psi = \pr(30) - \pr(10)$ & (all objects)                                      & $0.026$ & $16$ \\
$p^*$ & (all objects)                                                           & $0.043$ & $13$ \\
    & & & 		
		\end{tabular}
\end{table*}
\begin{table*}
\label{Tab:Tab5}
	\centering
	\caption{List of polarimetric parameters $h$, $h_{ABC}$ and $P_{\rm min}$ obtained for $15$
        asteroids of the \citet{ShevTed} list, having albedo larger than $0.08$, and the corresponding
        albedos computed using the calibrations based on these objects, only (see Table 1). $N_{tot}$ is the number of
        polarimetric observations available for each object. The last column gives,
        for a comparison, the albedo value listed by \citet{ShevTed}.}
		\begin{tabular}{rrccccccc}
		& & & \\
 \multicolumn{1}{c}{$N$} & \multicolumn{1}{c}{$N_{\rm tot}$} & \multicolumn{1}{c}{$h$} & \multicolumn{1}{c}{$h_{ABC}$} & 
 \multicolumn{1}{c}{$P_{\rm min}$} & \multicolumn{1}{c}{$p_V(h)$} & \multicolumn{1}{c}{$p_V(h_{ABC})$} &
 \multicolumn{1}{c}{$p_V(P_{\rm min})$} & \multicolumn{1}{c}{$p_V(ST)$} \\
		& & & \\
    1 & 33 & $0.2549 \pm 0.0041$ & $0.2467 \pm 0.0009$ & $-1.683 \pm 0.005$ & $0.102 \pm 0.010$ & $0.101 \pm 0.010$ & 
       $0.103 \pm 0.003$ & $0.094 \pm 0.006$ \\
    2 & 22 & $0.2223 \pm 0.0015$ & $0.2111 \pm 0.0004$ & $-1.413 \pm 0.017$ & $0.114 \pm 0.012$ & $0.114 \pm 0.012$ & 
       $0.120 \pm 0.003$ & $0.145 \pm 0.009$ \\
    3 & 26 & $0.1029 \pm 0.0039$ & $0.0949 \pm 0.0013$ & $-0.732 \pm 0.009$ & $0.210 \pm 0.027$ & $0.213 \pm 0.026$ & 
       $0.213 \pm 0.005$ & $0.187 \pm 0.006$ \\
    4 & 26 & $0.0691 \pm 0.0012$ & $0.0701 \pm 0.0005$ & $-0.558 \pm 0.005$ & $0.289 \pm 0.040$ & $0.270 \pm 0.035$ & 
       $0.270 \pm 0.009$ & $0.350 \pm 0.020$ \\
    8 & 28 & $0.1067 \pm 0.0016$ & $0.0950 \pm 0.0003$ & $-0.671 \pm 0.007$ & $0.204 \pm 0.026$ & $0.213 \pm 0.026$ & 
       $0.230 \pm 0.006$ & $0.197 \pm 0.012$ \\
   27 & 12 & \multicolumn{1}{c}{--} & $0.0582 \pm 0.0023$ & $-0.593 \pm 0.010$ &  \multicolumn{1}{c}{--} & $0.312 \pm 0.043$ & 
      $0.256 \pm 0.008$ & $0.298 \pm 0.045$ \\
   39 & 20 & $0.0745 \pm 0.0093$ & $0.0906 \pm 0.0009$ & $-0.712 \pm 0.013$ & $0.272 \pm 0.046$ & $0.221 \pm 0.027$ & 
      $0.218 \pm 0.006$ & $0.246 \pm 0.015$ \\
   51 & 20 & $0.2713 \pm 0.0040$ & $0.2644 \pm 0.0013$ & $-1.950 \pm 0.015$ & $0.097 \pm 0.010$ & $0.096 \pm 0.009$ & 
      $0.091 \pm 0.003$ & $0.097 \pm 0.006$ \\
   64 & 13 & $0.0434 \pm 0.0043$ & $0.0395 \pm 0.0014$ & $-0.323 \pm 0.014$ & $0.420 \pm 0.073$ & $0.422 \pm 0.063$ & 
      $0.434 \pm 0.027$ & $0.474 \pm 0.047$ \\
   78 &  5 &  \multicolumn{1}{c}{--} & $0.2394 \pm 0.0307$ & $-1.497 \pm 0.048$ &  \multicolumn{1}{c}{--} & $0.104 \pm 0.015$ &
      $0.114 \pm 0.004$ & $0.086 \pm 0.003$ \\
  124 &  5 & $0.0745 \pm 0.0041$ &  \multicolumn{1}{c}{--} &  \multicolumn{1}{c}{--} & $0.272 \pm 0.039$ &  \multicolumn{1}{c}{--} & 
       \multicolumn{1}{c}{--} &  $0.240 \pm 0.036$ \\
  129 & 10 & $0.0953 \pm 0.0077$ & $0.0928 \pm 0.0004$ & $-0.849 \pm 0.011$ & $0.224 \pm 0.032$ & $0.217 \pm 0.026$ & 
      $0.187 \pm 0.004$ & $0.183 \pm 0.018$ \\
  216 &  9 &  \multicolumn{1}{c}{--} & $0.1971 \pm 0.0041$ & $-1.028 \pm 0.011$ &  \multicolumn{1}{c}{--} & $0.121 \pm 0.012$ & 
      $0.159 \pm 0.003$ & $0.170 \pm 0.010$ \\
  230 & 15 & $0.1045 \pm 0.0047$ & $0.1089 \pm 0.0010$ & $-0.937 \pm 0.012$ & $0.208 \pm 0.027$ & $0.191 \pm 0.022$ & 
      $0.172 \pm 0.004$ & $0.192 \pm 0.019$ \\
  431 &  7 &  \multicolumn{1}{c}{--} & $0.2478 \pm 0.0023$ & $-1.360 \pm 0.024$ &  \multicolumn{1}{c}{--} & $0.101 \pm 0.010$ & 
      $0.124 \pm 0.003$ & $0.122 \pm 0.018$ \\
		\end{tabular}
\end{table*}

According to current evidence, we know that low albedo asteroids
exhibit a deeper value of $P_{\rm min}$ as well as a steeper linear
polarization slope over a large interval of phase angles. Asteroids
having increasingly higher albedos exhibit an opposite behavior
(increasingly shallower $P_{\rm min}$ and gentler $h$).  There are
also differences in the typical values of the inversion angle for
different classes of objects, as found, as an example, in the case of
the $F$ taxonomic class by \citet{Irina05}.  One can imagine many
different ways to attempt a new calibration of the albedo -
polarization relationship trying to exploit the above evidence and
make use of the overall morphology of the phase - polarization curve.
A first attempt in this direction was proposed by \citet{Masiero12}.

These authors proposed to use a new observable, they called $p\ast$
(\emph{p-star}), defined as a parameter of maximum polarimetric
variation, given by
\[p\ast = W_1 \log(h) + W_2 \log(P_{\rm min})
\]
where $h$ is, again, the classical polarimetric slope, and $W_1$ and
$W_2$ are two parameters whose values were found by \citet{Masiero12}
to be $W_1 = 0.79 \pm 0.02$ and $W_2 = 0.61 \pm 0.03$. The authors
used in their analysis a data-set of $177$ asteroids having an albedo
value estimated from thermal radiometry data produced by the WISE
mission, whereas polarimetric data were taken by the authors from the
literature. For $65$ asteroids of this sample, the authors derived a
value of $h$, while for $112$ objects they derived a value of $P_{\rm  min}$.  
In this way, \citet{Masiero12} derived a new calibration of
the albedo - polarization relationship based on WISE albedos and the
newly introduced $p\ast$ parameter. The relation they found was:
\[ \log(p_V) = C\ast_1\: p\!\ast + C\ast_2 \]
and they found for the $C\ast_1$ and $C\ast_2$ coefficients the values 
$C\ast_1 = -1.04 \pm 0.04$ and $C\ast_2 = -1.58 \pm 0.09$.

Since we use in our analysis a much smaller number of asteroids having
presumably more accurate albedo values not derived from thermal
radiometry, and we use a different set of polarimetric data including
a large number of previously unpublished observations, we have decided
to derive a new calibration of the albedo - $p\ast$ relation, using
the data at our disposal, while keeping the same definition of the
$p\ast$ parameter as given by \citet{Masiero12}. In particular, we
kept the values computed by \citet{Masiero12} for the $W_1$ and $W_2$
parameters, which were derived by the authors based on their own
analysis of polarimetric data.
We computed then the $p\ast$ parameter for our sample of asteroids
using $P_{\rm min}$ values already obtained from our best-fit of
Eq.$\:$\ref{Eqn:ABC} and polarimetric slopes $h$ derived as described
in Section$~$\ref{classical}. We had at disposal estimates for both $h$
and $P_{\rm min}$ for $13$ objects, only. The results of this exercise
are shown in Fig.$\:$\ref{Fig:pstarfit}, in which we also show the new
values that we found for the $C\ast_1$ and $C\ast_2$ parameters. 
In particular, we have:
\[C\ast_1 = -0.896 \pm 0.029\] 
and 
\[C\ast_2 = -1.457 \pm 0.018\]
The resulting fit is fairly good, and we confirm that $p\ast$ is 
another useful parameterfor albedo determination. 
The resulting RMS deviations, however, are higher
than those corresponding to the slope - albedo relation discussed
above, both using either $h$ or $h_{abc}$, as shown in Table 4.  We find 
a quite big discrepancy concerning the predicted albedo for (64) Angelina
and the corresponding \citet{ShevTed} value. In the region of
higher $p\ast$ values (right region of the plot), moreover, the
scatter of the albedos around the best-fit line is fairly high.

There are, of course, other possibilities to use polarimetric parameters
describing the overall morphology of the observed phase - polarization
curves. A very simple idea is to use some parameter built directly
from the obtained values of polarization \pr\ taken at very different
values of phase angle, possibly including both the negative and the
positive polarization branches.  In this paper, we introduce such a
parameter, that we call $\Psi$, and we define it as
	\[\Psi = \pr(30^\circ) - \pr(10^\circ)	\]
where the dependence of \pr\ upon the phase angle is assumed to be
given by Eq.$\:$\ref{Eqn:ABC}.  We derived therefore the $\Psi$
parameter for our sample of asteroids in the \citet{ShevTed} list
following the same criteria already described above in the case of our
computations of $P_{\rm min}$ and $h_{ABC}$. This corresponds to $20$
asteroids, listed in Table 2. For the purposes of calibration,  
again, we did not use data of
four asteroids having fewer than $10$ polarimetric measurements, 
which in Fig.$~$\ref{Fig:psialbloglogfit} are displayed
using different symbols. The Figure shows a log - log plot of $\Psi$
as a function of the \citet{ShevTed} albedo, 
according to the relation
\begin{displaymath}
\log(p_V) = C_{\psi1} \log(\Psi) + C_{\psi2}
\end{displaymath}
identical to the classical slope - albedo relation, but using
the new parameter $\Psi$ instead of the polarimetric slope $h$.
Fig.$~$\ref{Fig:psialbloglogfit} shows the best-fit
linear solution, and the resulting values of the $C\psi_1$ and $C\psi_2$
parameters, which are also listed, together with their
corresponding uncertainties, in Table 1.

The fit looks good, and this is confirmed by the resulting RMS deviations, 
which are found to be slightly 
lower than in the case of all other albedo - polarization relations
examined in this paper, including those obtained by excluding low-albedo
objects, as shown by the in Table 4. 
In Fig.$\:$\ref{Fig:diffalbpsi}, we plot the differences between 
the albedo values produced by our $\Psi$ - albedo calibration and the 
albedos given by \citet{ShevTed}. The albedo values obtained from the
$\Psi$ parameter tend to be always very close to the corresponding 
\citet{ShevTed} values, although the relative uncertainty tends
to be fairly high for the darkest asteroids.
 
With respect to the results obtained by using the polarimetric slope
as shown in previous Sections, 
it is encouraging to note that our $\Psi$ - albedo relation seems to 
fit nicely the \citet{ShevTed} albedos in the whole interval covered 
by the data. 
The albedo obtained for (64) Angelina seems to suggest that we
have no longer the problem of a possible overestimation of the albedo
of bright objects, which affected the calibration of the classical
slope - albedo law, as seen in Section \ref{classical}. 
We also note that the value of $C\psi_1 = -0.987 \pm 0.022$, obtained
in our calibration of the $\Psi$ - albedo relation, is also formally in 
agreement, within the uncertainty, with an even simpler
hyperbolic relation $p_V = K/\Psi$, with $K = 10^{C_2} \simeq 0.348$.

We find, again, some problems with asteroid (2) Pallas, for which an
albedo of $0.086 \pm 0.004$ is found by our $\Psi$ calibration, 
still much lower than the $0.145$ albedo indicated by
\citet{ShevTed}. The same is true for (216) Kleopatra, which has
a $\Psi$ value nearly identical to that of Pallas, corresponding to an 
albedo of $0.085 \pm 0.004$, much lower than the \citet{ShevTed} value of 
$0.170 \pm 0.010$. As opposite, for
(444) Gyptis we have problems of a remarkable overestimation of the
albedo, but in this case the \citet{ShevTed} albedo is extremely low,
$0.037 \pm 0.004$. Both Kleopatra and Gyptis are asteroids for which we
have fewer than $10$ polarimetric measurements. In the case of (444)
Gyptis, the phase - polarization curve is quite noisy, whereas this is
not the case for Kleopatra.

\section{Conclusions and future work}\label{conc}
The solution of the problem of determining the best possible
calibration of the relation between geometric albedo and polarization
properties for the asteroids is still an important task even in the
era of large thermal radiometry surveys, whose results are
valid in terms of statistical distribution among large samples of the
population, but can be strongly inaccurate for what concerns single
objects.

It is therefore very important to optimize the performances of the
polarimetric technique, as an effective tool to estimate the albedo of
the objects, with particularly important applications to the physical
characterization of newly-discovered, and potentially hazardous
near-Earth objects. We stress again that, once a reliable calibration
of the relation between albedo and polarimetric parameters is
available, the albedo values obtained by polarimetric data are not
affected at all by uncertainties due to poor knowledge of the absolute
magnitude of the objects, a relevant advantage over other possible
techniques.

In this paper we have carried out an extensive analysis, based on the
idea of using for calibration purposes a still limited number of
asteroids for which we can be reasonably confident to know the albedo
with good accuracy. We have used this sample to obtain a new
calibration of the classical slope - albedo law, with the polarimetric
slope being derived from available data using different possible
approaches.  We have also analyzed other possible relations, including
a new calibration of the classical $P_{\rm min}$ - albedo relation, the 
more recently proposed $p\ast$ - albedo relation \citep{Masiero12}, and 
a new relation based on the $\Psi$ polarimetric parameter, introduced for 
the first time in this paper. The resulting values of the various 
polarimetric parameters for $22$ asteroids considered in our analysis, 
and the corresponding values of albedo are given in Tables 2 and 3, 
respectively.

The extensive analysis presented in this paper produced a variety
of interesting results. For what concerns the ``classical'' slope -
albedo and $P_{\rm min}$ - albedo relations, we confirm that it is
difficult to find a calibration which can fit accurately objects
of all albedo classes. The results improve very much when the
asteroids having low albedo, below $0.08$, are not considered in the 
analysis. The presence of asteroids having significantly different 
albedo but nearly identical polarimetric slopes can even suggest that
a parabolic fit is more suited to represent the $h$ - albedo
relation, as shown in Section$~$\ref{lowpvslope} (see Fig.$~$\ref{Fig:Karri}).

The calibration of the $h$ albedo and $P_{\rm min}$ - albedo relations
that we obtained by excluding low-albedo asteroids from our analysis 
gives good results, with uncertainties within $\pm 20$\% for 
medium- and high-albedo objects. For asteroids exhibiting a steep
polarimetric slope, or having a polarimetric slope computed
on the basis of only a few observations, the classical $h$ - albedo relation  
can still be used, using the calibration coefficients found by considering
all the available calibration objects, since the expected errors
should be still acceptable. This is encouraging, because whenever 
there are at least a few measurements obtained at
phase angles larger than $14^\circ$, it is generally
possible to derive a value of the polarimetric slope $h$, and the 
resulting albedo values turn out to be reasonably accurate in
absolute terms, although
the relative error can be above $30$\% for low-albedo objects.
As opposite, the use of $P_{\rm min}$ 
should always be discouraged for asteroids having a deep negative 
polarization branch, as seen in Section \ref{Pmin}.

If one wants to avoid the complication of using different calibrations
of the slope - albedo relation for asteroids belonging 
to different albedo classes, other polarimetric parameters can 
also be adopted. In particular,
both $\Psi$ and $p\ast$ require to have at disposal phase - polarization 
curves of a sufficiently good quality to be fit by means of an 
exponential - linear representation (Eq.$\:$\ref{Eqn:ABC}).  
This cannot be done
when the number of polarimetric measurements is too small, or the
observations are concentrated over a too limited interval of phase
angles.
While the uncertainty of albedo determinations based on the
$p\ast$ parameter seems to be reasonable but not really negligible,
the most accurate albedo determinations can be
obtained when the $\Psi$ parameter can be reliably determined 
from the available data. The advantage of using $\Psi$ is that
this parameter seems to be suited to give accurate values of
albedo for both bright and dark asteroids. 

Of course, there are some {\em caveats} to be taken into account.
For instance, there are asteroids, like the so-called Barbarians, 
which exhibit peculiar phase - polarization curves 
\citep[see][]{Cellinoetal06}. The determination of the geometric albedo for
Barbarian asteroids is a problem, because any derivation of the
albedo using relations valid for the rest of the population may be
misleading. However, it is also true that more data are needed to
better understand the situation.

In fact, the major problem of asteroid polarimetry today is certainly
a serious lack of data. Further progress in this field will require
the use of dedicated telescopes, to fill the gap with the amount of
information already available from other observing techniques.

Polarimetry has been so far
severely under-appreciated as an essential tool for physical
characterization of asteroids and also other classes of small bodies
in the Solar System.  The present paper summarizes the current state
of the art for what concerns the calibration of the relation between
polarimetric properties and albedo, and makes some further steps
forward, with the introduction of the $\Psi$ parameter, which
seems to be a new useful tool to obtain reliable values of asteroid 
albedos. 

In a separate paper, to be submitted soon for publication, 
we exploit the results of the present analysis
to derive the albedos of a fairly large number of objects for which
we lack a \citet{ShevTed} determination, and we analyze the distributions
of other parameters that characterize the phase - polarization curves
of main belt asteroids. 

\section*{Acknowledgements} 
We thank the Referee Dr.$~$K.$~$ Muinonen for his important comments
and suggestions that led to substantial improvements of this paper. AC
was partly supported by funds of the PRIN INAF 2011. AC and SB
acknowledge support from COST Action MP1104 “Polarimetry as a tool to
study the solar system and beyond” through funding granted for Short
Terms Scientific Missions and participation to meetings.  RGH
gratefully acknowledges financial support by CONICET through PIP
114-201101-00358.

\label{lastpage}

\begin{thebibliography}{99}

\bibitem[Bagnulo et al.(1995)]{Bagnuloetal1995} Bagnulo, S., Landi Degl'Innocenti, E., Landolfi, M.,
\& Leroy, J.-L. 1995, A\&A, 295, 459

\bibitem[Belskaya et al.(2005)]{Irina05} Belskaya, I. N., Shkuratov, Tu., Efimov, Yu., et al. 
2005, Icarus, 178, 213

\bibitem[Bevington(1969)]{Bevington} Bevington P.R. 1969, 
Data reduction and Error analysis for the physical sciences, Mc Graw-Hill Book Company.

\bibitem[Ca\~nada-Assandri et al.(2012)]{Assandrietal2012} Ca\~nada-Assandri, M., Gil-Hutton, R.
\& Benavidez, P. 2012, A\&A, 542, A11

\bibitem[Cellino et al.(1999)]{Cellinoetal99}Cellino, A., Gil-Hutton, R.,
Tedesco, E.F., Di Martino, M. \& Brunini, A. 1999, Icarus, 138, 129

\bibitem[Cellino et al.(2006)]{Cellinoetal06}Cellino, A., Belskaya, I.N., Bendjoya, Ph., Di Martino, M.,
Gil-Hutton, R., Muinonen, K., \& Tedesco, E.F. 2006, Icarus, 180, 565

\bibitem[Cellino et al.(2012)]{Cellinoetal2012}Cellino, A., Gil-Hutton, R., Dell'Oro, A., Bendjoya, Ph.,
Ca\~nada-Assandri, M. \& Di Martino, M. 2012, JQSRT, 18, 2552

\bibitem[Cellino et al.(2015)]{CellinoDawn}Cellino, A., Ammannito, E., Magni, G., Gil-Hutton, R., 
Belskaya, I.N., Tedesco, E.F., \& De Sanctis, C. 2015, MNRAS (submitted)

\bibitem[De Leon et al.(2012)]{DeLeonetal12} De Leon, J., Pinilla-Alonso, N., Campins, H., Licandro, X. \&
Marzo, G.A. 2012, Icarus, 218, 196

\bibitem[Gil-Hutton and Ca\~nada-Assandri(2011)] {RGH11} Gil-Hutton, R. \& Ca\~nada-Assandri, M. 2011, 
A\&A, 529, A86

\bibitem[Gil-Hutton and Ca\~nada-Assandri(2012)] {RGH12} Gil-Hutton, R. \& Ca\~nada-Assandri, M. 2012,
A\&A, 539, A115

\bibitem[Gil Hutton et al.(2014)]{Giletal14} Gil Hutton, R., Cellino, A. \& Bendjoya, Ph. 2014,
A\&A, 569, A122

\bibitem[Kaasalainen et al.(2003)]{Shennaetal03} Kaasalainen, S., Piironen, J., Kaasalainen, M., Harris, A.W.,
Muinonen, K, \& Cellino, A. 2003, Icarus, 161, 34

\bibitem[Lupishko and Mohamed(1996)]{LupMoh}Lupishko, D.F. \& Mohamed, R.A. 1996, Icarus, 119, 209

\bibitem[Masiero et al.(2011)]{MasieroWISE}Masiero, J.R., Mainzer, A.K., Grav., T., et al. 2011, ApJ, 741, 66

\bibitem[Masiero et al.(2012)]{Masiero12}Masiero, J.R., Mainzer, A.K., Grav, T., Bauer, J.M.,  Wright, E.L.,
Mc Millan, R.S., Tholen, D.J. \& Blain, A.W. 2012, ApJ, 749, A104

\bibitem[Morrison and Lebofsky(1979)]{MorrisonLebofsky79}Morrison, D. \& Lebofsky, L. 1979. 
In \emph{Asteroids} (T. Gehrels, Ed.), pp.184-205. Univ. of Arizona Press, Tucson, AZ 

\bibitem[Muinonen et al.(2009)]{Muinonenetal09}Muinonen, K., Penttila, A., Cellino, A., Belskaya, I. N., 
Delbo, M., Levasseur-Regourd, A.-C., \& Tedesco, E.F. 2009, Meteoritics and Planet. Sci., 44, 1937

\bibitem[Muinonen et al.(2010)]{HG1G2}Muinonen, K., Belskaya, I.N., Cellino, A., Delb\`o, M., 
Levasseur-Regourd, A.C., Penttila, A. \& Tedesco, E.F. 2010, Icarus, 209, 542

\bibitem[Penttila et al.(2005)]{Penttilaetal2005} Penttila, A., Lumme, K., Hadamcik, E., \&
Levasseur-Regourd, A.C. 2005, A\&A, 432, 1081

\bibitem[Russel(1916)]{Russel16}Russel, H.N. 1916, Ap.J., 43, 173-195

\bibitem[Shevchenko and Tedesco(2006)]{ShevTed} Shevchenko, V.G., \& Tedesco, E.F. 2006, Icarus, 184, 
211-220

\bibitem[Tedesco and Veeder(1992)]{IMPS} Tedesco, E.F., \& Veeder G.J. 1992.
In \emph{IRAS Minor Planet Survey} (E.F. Tedesco, G.J.Veeder, J.W. Fowler, J.R. Chillemi, Eds.), 
pp. 243-285. Phillips Laboratory Final Report PL-TR-92-2049. Hanscom Air Force Base, MA

\bibitem[Tedesco et al.(2002)]{SIMPS} Tedesco, E.F., Noah, P.V., Noah, M. \& Price, S.D. 2002, AJ, 123, 1056

\bibitem[Tholen and Barucci (1989)]{TholenBar} Tholen, D. J., \& Barucci, M. A. 1989. In
\emph{Asteroids II} (R.P. Binzel, T. Gehrels, M.S. Matthews, Eds.) pp298-315. University of Arizona Press, Tucson

\bibitem[Umov(1905)]{Umov1905}Umov, N.\ 1905, Physik. Z., 6, 674

\bibitem[Usui et al.(2013)]{Usui13}Usui, F., T. Kasuga, Hasegawa, S., Ishiguro, M., Kuroda, D., M\"uller, T.G.,
Ootsubo, T. \& Matsuhara, H. 2013, ApJ, 762, article id. 56.

\bibitem[Zellner et al.(1974)]{Zellneretal74} Zellner, B., Gehrels, T. \& Gradie, J. 1974, AJ, 79, 1100

\bibitem[Zellner and Gradie(1976)]{ZellnerGradie76}Zellner, B., \& Gradie, J. 1976, AJ, 81, 262

\bibitem[Zellner et al.(1977)]{Zellneretal77}Zellner, B., Leake, M.,
Lebertre, T., Duseaux, M., \& Dollfus, A. 1977, Proc. Lunar Sci. Conf. 8th, p. 1091
\end{thebibliography}
\end{document}